\documentclass[sigplan,nonacm]{acmart}
\renewcommand\footnotetextcopyrightpermission{}
\setcopyright{none}
\usepackage{xspace}
\usepackage{subcaption}
\usepackage{enumitem}
\usepackage[]{hyperref}
\usepackage[normalem]{ulem}
\usepackage{tikz}
\usetikzlibrary{arrows.meta}

\renewcommand{\emph}[1]{\textit{#1}}
\newcommand{\sys}{SnapShare\xspace} % obfuscate

\ifdefined\RELEASE
  \newcommand{\ignore}[1]{}
  \newcommand{\fixme}[1]{}
  \newcommand{\dmi}[1]{}
  \newcommand{\leo}[1]{#1}
  \newcommand{\kev}[1]{}
  \newcommand{\myai}[1]{}
  \newcommand{\rod}[1]{}
  
  \newcommand{\TODO}[1]{}

\else
  \newcommand{\ignore}[1]{}
  \newcommand{\fixme}[1]{{\textcolor{red}{[~FIXME:~#1~]}}}
  \newcommand{\dmi}[1]{{\textcolor{blue}{[~D:~#1~]}}}
  \newcommand{\leo}[1]{{\textcolor{teal}{[~L:~#1~]}}}
  \newcommand{\kev}[1]{{\textcolor{brown}{[~K:~#1~]}}}
  
  \newcommand{\myai}[1]{{\textcolor{olive}{[~AI:~#1~]}}}
  \newcommand{\rod}[1]{{\textcolor{green}{[~R:~#1~]}}}
  
  \newcommand{\TODO}[1]{{\textcolor{red}{TODO:~#1}}}
  
\fi

\begin{document}
%-------------------------------------------------------------------------------

%don't want date printed
\date{}

% \title{\sys: Safe and Transparent Snapshot Sharing at the Cloud Scale}
% \title{\sys: Provenance-Aware Snapshot Deduplication for Serverless Clouds}
\title[\sys]{\sys: Isolation-Preserving Deduplication for Serverless Snapshots}
% \title{\sys: Cluster-Wide Snapshot Sharing for Faster Serverless Cold Starts}

\author{Leonid Kondrashov}
\affiliation{%
  \institution{NTU Singapore}
  \country{Singapore}
}
\author{Likun Zhang}
\affiliation{%
  \institution{NTU Singapore}
  \country{Singapore}
}
\author{Andr\'{e} Filipe do Pilar de Jesus}
\affiliation{%
  \institution{Instituto Superior T\'{e}cnico, University of Lisbon}
  \country{Portugal}
}
\author{Hans Pistor}
\affiliation{%
  \institution{AWS}
  \country{United States}
}
\author{Marco Cali}
\affiliation{%
  \institution{AWS}
  \country{United Kingdom}
}
\author{Riccardo Mancini}
\affiliation{%
  \institution{AWS}
  \country{United Kingdom}
}
\author{Rodrigo Bruno}
\affiliation{%
  \institution{INESC-ID, Instituto Superior T\'{e}cnico, University of Lisbon}
  \country{Portugal}
}
\author{Dmitrii Ustiugov}
\affiliation{%
  \institution{NTU Singapore}
  \country{Singapore}
}

% \dmi{}\leo{}\kev{}\rod{}

%-------------------------------------------------------------------------------
\begin{abstract}
%-------------------------------------------------------------------------------
% The efficiency of serverless computing fundamentally depends on rapid function startup, a challenge increasingly addressed by restoring from virtual machine snapshots. However, production platforms like AWS Lambda isolate these snapshots per-tenant using convergent encryption to prevent timing-channel attacks. \leo{this is the opposite, the convergent encryption opens the timing channel. the strict no-sharing is what fixes it} While secure, strict isolation precludes the deduplication of common infrastructure components, wasting per-worker cache capacity and exacerbating cold-start latency through inefficient remote fetching. Our characterization reveals that between 50\% \lik{11\%} and 65\% of a typical snapshot's working set consists of non-sensitive, shareable components.  \lik{or using on average 50\%}
% \leo{updated to match the rest of the paper}
The efficiency of serverless computing fundamentally depends on rapid function startup, a challenge increasingly addressed by restoring from virtual machine snapshots. However, because content deduplication techniques can open timing-channel attacks that leak sensitive tenant data, production platforms like AWS Lambda instead enforce strict, zero-sharing isolation between snapshots of different tenants. While secure, this isolation precludes the deduplication of common infrastructure components, wasting per-worker cache capacity and exacerbating cold-start latency through inefficient remote fetching. Our characterization reveals that non-sensitive, shareable infrastructure accounts for roughly 70\% of a typical snapshot's contents and, on average, about 50\% of its working set.

We present \sys, a cluster-wide snapshot management system that co-designs storage and caching to exploit this commonality without compromising sensitive data isolation. \sys classifies a snapshot page as shareable only when its full content matches that of a page from a provider-approved source; any non-matching content is treated as isolated private content. Common infrastructure components are selectively deduplicated and pinned across the cluster's worker nodes, while private data remains strictly isolated and is eagerly fetched as part of the tracked working set, falling back to on-demand retrieval only for pages that fall outside it.
%By integrating snapshot management with cache-aware scheduling, \dmi{this is stale, we removed sched-cache co-design, now it's cache-stor co-design}
%\sys drastically reduces data movement on the critical path. 
Implemented as an extension to Knative, \sys reduces end-to-end tail request slowdowns by 10–29\% compared to traditional working-set and chunk-based platforms. It also reduces remote storage requirements by 39\% and local cache footprints by 38\%, reducing network retrieval payloads and bounding worst-case fetch delays by 10$\times$.
% \TODO{update the numbers}
%\sys reduces cold-start retrieval latency by up to 9$\times$ compared to chunk-based distribution systems like AWS SnapStart. Furthermore, it improves storage and cache efficiency by 40\% to 50\% --- matching the performance of insecure full-deduplication systems --- and maintains sub-100ms fetch times even under heavily constrained (1 Gbps) network bandwidth.
%\dmi{the quantitative part of the abstract is stale/irrelevant: please take a look at \S\ref{sec:eval:summary} for the latest. Particularly, 1Gbps is off because we measure most of our experiments with 10Gbps limit, not 1Gbps; intro's contributions seem up to date}
\end{abstract}

\maketitle

% !TEX root = preview-ai.tex
% Tracked review edits: main.tex displays myai additions and sout deletions.
% preview-ai.tex hides myai blocks and is NOT an accepted-changes preview.
% Architecture and policy artwork retained pending redraw; the compact source
% breakdown summarizes reported approximate fractions, not new measurements.
%
% CURRENT REVIEW: numerical edits below use the latest 17-function compressed
% evaluation (39.4% storage / 38.4% cache versus Sabre). Older notes below are
% historical and must not override the current evaluation.
% HISTORICAL NUMBERS TO RECONCILE (no new measurements assumed):
% - Characterization: about 50% of working sets match infrastructure sources;
%   the full-snapshot fraction is 65%, with a newer 70% annotation. Use
%   "more than half" for full snapshots without conflating the denominators.
% - Storage/cache: retain the current evaluation's 50% / 40% footprint savings.
% - Tail slowdown: use the newer 10% (WS) / 29% (Chunks) annotations in
%   sections/6-evaluation.tex, not the older 18--43% headline. These are
%   reductions in the geometric mean of per-function P99 request slowdowns.
% - "Within 5% of Full Dedup" does not establish "5% residual duplication";
%   keep that storage comparison qualitative until its denominator is reconciled.
% - Full Dedup's additional 5% fetch-size / <1% restoration-time reductions,
%   and the 5% cache penalty without image sharing, are measured results from
%   the supplied EuroSys rebuttal. They are separate from the updated trace
%   experiment, and must not be presented as its P99 slowdown gaps.

\section{Introduction}

\begin{figure}[t]
    \centering
    \includegraphics[width=0.95\linewidth]{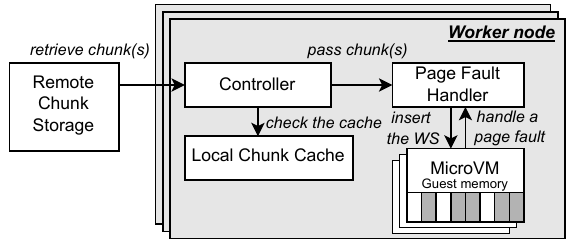}
    \caption[Snapshot restoration through a worker-local cache.]{Snapshot restoration through a worker-local cache. A cold start fetches missing state from remote storage; duplicate copies of common infrastructure consume capacity in both tiers. }
    \label{fig:ai:baseline_arch}
\end{figure}

Serverless elasticity is limited by how fast function instances are created during workload changes~\cite{datadog}. Platforms such as AWS Lambda~\cite{aws-snap-serv} accelerate function instance cold starts by restoring virtual machines from snapshots. As Figure~\ref{fig:ai:baseline_arch} illustrates, workers cache snapshot chunks locally and fetch missing ones from remote storage. The capacity of this cache determines how much state can be reused during a cold start. However, snapshots of different functions contain duplicate copies of common infrastructure data found in popular container and base VM images. Duplicate snapshot data wastes remote storage and displaces useful state from worker caches, forcing more cold starts and remote state retrieval.

% REDRAW BRIEF, Fig. 1: retain remote storage, worker cache, and guest memory;
% make the local-hit / remote-miss paths explicit. Show repeated infrastructure
% copies consuming cache slots and remote storage, with consistent colors in
% Fig. 3. Avoid expanding this introductory figure into a controller diagram.

\begin{figure}[t]
    \centering
    \begin{subfigure}{0.95\linewidth}
        \centering
        \includegraphics[width=\linewidth]{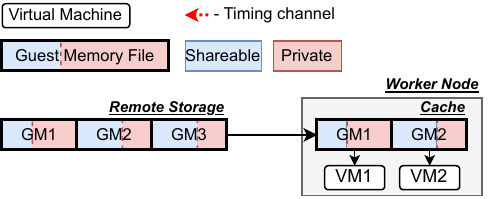}
        \caption{No cross-tenant sharing, as in AWS Lambda~\cite{agache:firecracker}, REAP~\cite{ustiugov:benchmarking}.}
        \label{fig:ai:duplication_ws}
    \end{subfigure}

    \smallskip
    \begin{subfigure}{0.95\linewidth}
        \centering
        \includegraphics[width=\linewidth]{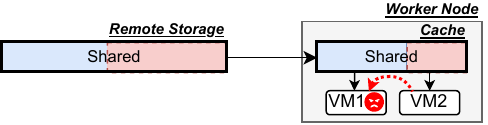}
        \caption{Unrestricted deduplication, as in Medes~\cite{saxena:memory}, SnapStore~\cite{panda:snapstore}.}
        \label{fig:ai:duplication_full_dedup}
    \end{subfigure}

    \smallskip
    \begin{subfigure}{0.95\linewidth}
        \centering
        \includegraphics[width=\linewidth]{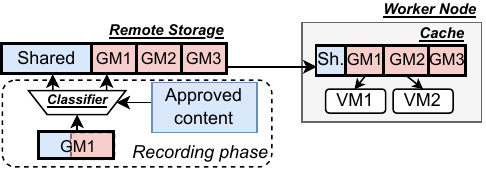}
        \caption{\sys.}
        \label{fig:ai:duplication_splitsnap}
    \end{subfigure}

    \caption[Snapshot sharing policies and their capacity cost.]{Snapshot sharing policies and their capacity cost. (a) Separate copies duplicate common infrastructure. (b) Unrestricted deduplication also shares private content. (c) \sys deduplicates approved reference content while retaining separate private content. 
    % \dmi{Delimit the legend, placing it on top (currently, it is inside all these diagrams);}
    % \dmi{IMPORTANT: "VM" is confusing bc VMs on worker nodes are different from the snaps inside storage and cache, let's call it "GM" instead and specify in the legend that this stands for guest memory contents}
    % \dmi{need to shrink the fig vertically; e.g., can reduce the "VM"s inside worker nodes}
    % \dmi{make VM borders thicker so it's clear that blue and red are inside these boxes; make the border bw red and blue dashed lines}
    % }
    % \dmi{increase fonts for Rem stor, Node, Cache; minimize whitespace}
    % \dmi{add refs in the a) and b) captions to Firecracker, Medes, etc}
    % \rod{The introduction mentions Shared and Private, not Shared and Sensitive as in (a).}
    }
    \label{fig:ai:duplication_comparison}
\end{figure}

%\textbf{The barrier is reconciling snapshot isolation with efficient caching.} 
Production platforms store VM snapshots encrypted in remote storage~\cite{aws-snap-serv}, but encryption alone does not prevent deduplication-induced disclosure: reusing a victim's cached content can change an attacker's retrieval time, revealing the presence of sensitive state or cryptographic keys~\cite{suzaki:memory-disclosure,xiao:memory-dedup-security,bosman:dedup,schwarzl:remote-dedup}. Production platforms thus take a conservative isolation approach that forces separate copies across security domains~\cite{firecracker_docs,private_comm}, treating all snapshot state (including public infrastructure state) as private state (Figure~\ref{fig:ai:duplication_ws}). While container-based platforms already share infrastructure artifacts to accelerate cold starts~\cite{brooker:on-demand}, guest memory snapshots complicate this opportunity by intermixing infrastructure bytes with customer-specific run-time state. Prior snapshot-deduplication systems, including SnapStore~\cite{panda:snapstore} and Medes~\cite{saxena:memory} depicted in Figure~\ref{fig:ai:duplication_full_dedup}, demonstrate the snapshot deduplication potential but lack the critical capability to distinguish the state that is safe to share from the sensitive data. Instead, boths systems treat all content as shareable, opening up surface for timing attacks. 
% However, previous work treats all content as shareable (Figure~\ref{fig:ai:duplication_full_dedup}). %The missing capability is to identify eligible content inside guest memory without extending sharing to private state.

%\textbf{Characterization: infrastructure redundancy persists on the cold-start path.} 
% Original review comments retained for validation (before the 5999e07 edits):
% \rod{We need to explain this experiment. Was this SeBS? Figure 2 adds very little value. It only offers two numbers that we also mention in the text. I would drop it.}

\begin{figure}[t]
    \centering
    \includegraphics[width=\linewidth]{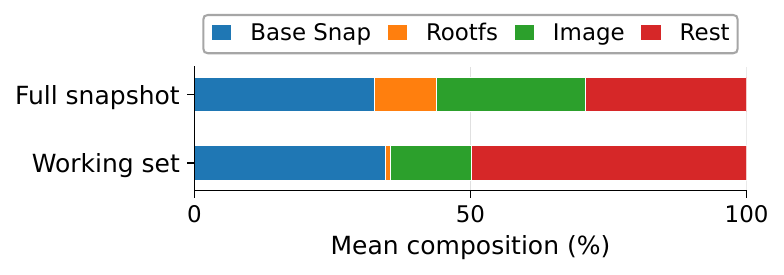}
    \vspace{-25pt}
    \caption{Mean source composition of full snapshots and cold-start working sets. Base-snapshot, rootfs, and container-image matches account for roughly 70\% of nonzero full-snapshot content and 50\% of working sets. Source composition is distinct from the duplicate-storage fraction in Figure~\ref{fig:back:duplication_storage}.}
    \Description{Stacked bars distinguish base snapshot, rootfs, container image, and unmatched content. Infrastructure matches occupy about seventy percent of full snapshots and half of working sets.}
    \label{fig:back:source_composition}
    \label{fig:ai:source_breakdown} % Compatibility for the struck-out old reference.
    \vspace{-10pt}
\end{figure}

% \dmi{rewritten, check}
\textbf{Key Insights.}
To quantify the shareable infrastructure snapshot state, we compare and classify the snapshot pages across three reference sources accessed during execution, when running a mix of 17 vSwarm~\cite{vswarm} functions written in Go, Python, and Node.js (methodology detailed in \S\ref{sec:methodology}). We differentiate all remaining pages (\textit{Rest}) and conservatively attribute them to the application itself. These sources include the pages found in the booted guest-OS (base) snapshot, VM root filesystem, and function container image. Also, we record the \emph{working set}, i.e., the pages accessed after restoration and initial request execution.
% \dmi{did you mean accessed after restoration?}
Figure~\ref{fig:back:source_composition} shows that \textbf{infrastructure matches account for roughly 70\% of a nonzero snapshot's content and 50\% of the working set} in a mix of workloads. These matches identify potential sharing opportunities, not automatic permission to share or bytes saved. Our detailed breakdown (\S\ref{sec:characterization}) shows that base-snapshot and rootfs matches recur ubiquitously across functions, whereas image matches recur mainly among functions using the same language runtime.

% \begin{figure}[t]
%     \centering
%     \input{pics/intro-source-breakdown-ai.tex}
%     \caption{Infrastructure content remains substantial in cold-start working sets. Approximate source-match fractions summarized from Figure~\ref{fig:bench_breakdown}; full snapshots exclude zero pages. Matching content becomes shareable only with provider approval.}
%     \Description{Two horizontal stacked bars: approximately two thirds of full-snapshot content and one half of working-set content match infrastructure references; the remainder is unmatched.}
%     \label{fig:ai:source_breakdown}
% \end{figure}

% Resolved: aggregate source-composition discussion incorporated into the characterization paragraph above; no duplicate paragraph needed.
% \TODO{Much of the opportunity comes from common infrastructure. Figure~\ref{fig:back:source_composition} breaks down snapshot content by matches to the base guest-OS snapshot, root filesystem, and container image; \emph{Rest} denotes unmatched content. These account for roughly 70\% of nonzero full-snapshot content and half of working sets. This source-match fraction is not the fraction of bytes saved by deduplication, which depends on recurrence across snapshots. The challenge is to maximize deduplication while preserving the isolation of private content.}

% REDRAW BRIEF, Fig. 3: use the same two tenants and logical page contents in
% all panels. Distinguish approved reference content from private content;
% show that even equal private pages retain separate copies in SplitSnap.
% The comparison should expose storage/cache duplication, not imply shared
% physical guest RAM. Preserve readable labels at one-column width.

\textbf{Key contribution: page attribution that enables selective deduplication.} 
% Original review comments retained for validation (before the 5999e07 edits):
% \rod{Not clear what this sentence means.}
We present \textit{\sys},\footnote{We will release the code and the toolchain upon publication.} a snapshot management system that co-designs remote storage and worker-local caching around a page-level sharing boundary. When a function is deployed, a snapshot is taken and an offline classifier compares the contents of each 4~KB snapshot page against provider-approved references. \textbf{Only complete matches are Shared; every other page is Private} (Figure~\ref{fig:ai:duplication_splitsnap}). A page loaded from an approved image but modified to contain application state stays private.
% unless its entire contents match an approved reference.
The \sys classifier checks what the page contains at snapshot time; it does not assume that all pages loaded from an approved image remain shareable. 
% \sout{Approval applies to snapshot bytes, not all memory associated with an image; this is content attribution, not execution-history tracking.}
Independently, deployment-time \texttt{userfaultfd} profiling identifies the \emph{Working Set} for eager retrieval and the remaining \emph{Lazy} pages.

%\textbf{Attribution exposes a useful, operator-controlled sharing boundary.} 
\textbf{Attribution exposes a useful, operator-controlled sharing boundary.}
% Original review comments retained for validation (before the 5999e07 edits):
% \rod{What characterization?}
% \rod{I don't understand this. This seems like container-image do not offer opportunity for deduplication.}
% \leo{it indeed does, it makes a small difference}
% \dmi{rewritten}
The trusted operator approves reference content and acceptable disclosure of its use; private pages remain tenant-separated even when equal (\S\ref{sec:back:security}). Operators can exclude libraries or images, including in response to published CVEs. The source-recurrence study (\S\ref{sec:characterization}) explains why this remains useful: common base-snapshot and rootfs pages supply much of the reuse. With container-image sharing disabled, cache savings relative to the best safe baseline, Sabre~\cite{lazarev:sabre}, decline from 38\% to 31\% (\S\ref{sec:eval:storage}). Image sharing therefore provides an additional benefit, but is not required to obtain substantial savings. 
% \sout{The trusted operator approves reference content and acceptable disclosure of its use; private pages remain separate across tenants even when equal (\S\ref{sec:back:security}). Operators can exclude particular libraries or images, including in response to published CVEs. The characterization  explains why conservative policies remain useful: common base-snapshot and rootfs pages supply much of the reuse. In our measured stricter-policy comparison, disabling container-image sharing increases local cache requirements by only 5\% .}}

%\textbf{Design: selective sharing with tenant-local private caching.} 
\sys stores approved shared components once across snapshots and pins common components in worker caches according to operator policy. Private components remain cacheable for reuse by their owning function, but retain tenant and function-version identity rather than being shared across tenants. Both storage and caches distinguish shared and private working-set files from shared and private lazy pages. During a cold start, the controller reuses resident shared components and eagerly fetches the missing private working set; the page-fault handler serves accesses outside the recorded working set on demand at 4~KB granularity. This avoids placing every initial page fetch on the critical path while preserving the sharing boundary. \sys builds on working-set coalescing and eager restoration from REAP~\cite{ustiugov:benchmarking} and FaaSnap~\cite{ao:faasnap}. Selective sharing adds capacity for more private working sets and avoids repeated infrastructure fetches on misses; the smaller payload also complements compression techniques such as Sabre~\cite{lazarev:sabre}.

%\textbf{Evaluation: substantial savings, close to unsafe full deduplication.}
% Original review comments retained for validation (before the 5999e07 edits):
% \leo{it is compressed}
% \rod{At this point here this baseline was not explained.}
% \rod{Same for this one. Shouldn't we mention REAP?}
% \leo{i would say we should use the same sabre (as we call it in the eval)}
% \rod{Same here.}
% \leo{can put chunk-based distribution}
% \leo{sabre}
% \dmi{rewritten, check}
We implement \sys atop vHive~\cite{ustiugov:benchmarking} and Firecracker~\cite{agache:firecracker}, evaluating 17 vSwarm~\cite{vswarm} functions and a workload driven by 700 Azure Functions~\cite{shahrad:serverless} trace profiles. All evaluated systems use streaming compression. Relative to the tenant-isolated Sabre~\cite{lazarev:sabre} baseline, which combines coalesced working-set retrieval with streaming compression, \sys reduces remote storage by 39\% and cache footprint by 38\% (\S\ref{sec:eval:storage}). \sys reduces the geometric mean of per-function P99 request slowdowns by 29\% versus the chunk-based baseline representative of a production AWS Lambda baseline~\cite{aws-snapstart-optimization} and 10\% versus Sabre (\S\ref{sec:eval:e2e}). Its storage and cache footprints remain within 6\% and 3\% of the idealized unsafe, fully deduplicated baseline, which also shares all snapshot content.
% \sout{We implement \sys atop vHive~\cite{ustiugov:benchmarking} and Firecracker~\cite{agache:firecracker}, evaluating vSwarm~\cite{vswarm} functions with 700 Azure Functions~\cite{shahrad:serverless} trace profiles. \sys halves uncompressed  remote storage relative to tenant-isolated Pages  and reduces cache footprint by 40\% relative to tenant-isolated WS   (\S\ref{sec:eval:storage}). With 4~GB worker caches, it lowers the geometric mean of per-function P99 request slowdowns by 29\% versus Chunks , the safe production-inspired baseline, and 10\% versus WS , the safe academic baseline incorporating techniques from REAP~\cite{ustiugov:benchmarking} and FaaSnap~\cite{ao:faasnap} (\S\ref{sec:eval:e2e}). In a separate restore-path comparison, the idealized unsafe Full Dedup oracle reduces fetch size by only another 5\% and restoration time by less than 1\%: deduplicating private content buys little additional restoration performance.}}

Our contributions include:
\begin{itemize}[leftmargin=.20cm,itemindent=5pt,labelwidth=2pt,labelsep=.2cm,align=left]
\setlength\itemsep{0pt}
    \item A characterization study showing that common infrastructure accounts for approximately half of cold-start working sets, exposing substantial reuse opportunities even when private application state cannot be shared.

    \item A page-attribution method that enables policy-controlled sharing through complete matches to approved reference content, with all other pages private by default. 
    %\leo{the data is correct but sounds weird to be here, not in the next point}
    % \dmi{check the numbers}
    % \sout{Excluding container-image sharing increases local cache requirements by only 5\% in our stricter-policy comparison.}}

    \item A classification-driven storage and cache co-design delivering 39\% lower remote-storage requirements and 38\% smaller cache footprints than Sabre, with 10--29\% lower aggregate tail request slowdown than safe baselines. Storage and cache footprints remain within 6\% and 3\% of unsafe fully deduplicated baseline, respectively. Even without container-image sharing, the system reduces cache footprint by 31\% relative to Sabre~\cite{lazarev:sabre}.
    
    % \dmi{I guess the data below is stale, check pls}
    % \sout{A classification-driven storage and cache design delivering 50\% lower uncompressed remote storage requirements, 40\% smaller working-set cache footprints, and 10--29\% lower aggregate tail request slowdown than safe baselines. In a separate restore-path comparison, unsafe Full Dedup offers only another 5\% reduction in fetch size and less than 1\% in restoration time.}

\end{itemize}

% \input{sections/1-introduction}

% !TEX root = preview-ai.tex
% Combined replacement for sections/2-background and sections/3-characterization
% in the alternate preview. Uses existing experiments and figure assets.
% The supplied aggregate plot shows roughly 70% full-snapshot infrastructure
% matches, distinct from the 43.9% duplicate-storage fraction and the 50%
% working-set source-match fraction. Cache hit rates are read approximately.

\section{Background and Motivation}
\label{sec:back}

Snapshot restoration trades repeated initialization for state retrieval. Its efficiency depends on how much state workers can reuse locally and how much they must fetch remotely. We first explain why duplicated guest memory makes this trade-off expensive and quantify the opportunity to recover capacity through selective deduplication. Then we identify the security constraints and define our security model.

\subsection{Duplication and Local Cache Efficiency}
\label{sec:back:restoration}

Serverless platforms create new instances as demand grows. Functions are commonly deployed as container images and executed in lightweight VMs such as Firecracker~\cite{agache:firecracker}, with image contents made available through the VM's filesystem~\cite{firecracker-container,brooker:on-demand}. A snapshot captures initialized guest memory and virtual disk state, avoiding repeated OS and application startup. As Figure~\ref{fig:ai:baseline_arch} shows, the selected worker retrieves snapshot content through its local cache and fetches missing state remotely. 
Local cache contains snapshot content organized in chunks, pages, or working sets~\cite{ustiugov:benchmarking,ao:faasnap}.\footnote{A working set comprises the pages accessed while restoring an instance and serving its initial execution; it is a subset of the full snapshot.}
% \sout{The \emph{cold-start working set} comprises the pages accessed while restoring an instance and serving its initial execution; it is a subset of the full snapshot. }

% \sout{Production-inspired restoration} 
In production platforms, upon a cold start, the system
% \TODO{ref to blog} \leo{already cited in the sentence}
retrieves state in chunks~\cite{aws-snapstart-optimization,aws_lambda_arch}, potentially transferring unaccessed pages alongside accessed ones. REAP~\cite{ustiugov:benchmarking} and FaaSnap~\cite{ao:faasnap} reduce this overhead by recording and coalescing \emph{working-set} pages for eager retrieval. However, each tenant's working set still contains duplicate copies of the same infrastructure. Providers also limit snapshot-retrieval bandwidth to leave capacity for application traffic~\cite{private_comm}, making avoidable remote fetches particularly undesirable.

\begin{figure}[t]
    \centering
    \begin{subfigure}[t]{0.48\linewidth}
        \centering
        \includegraphics[width=\linewidth]{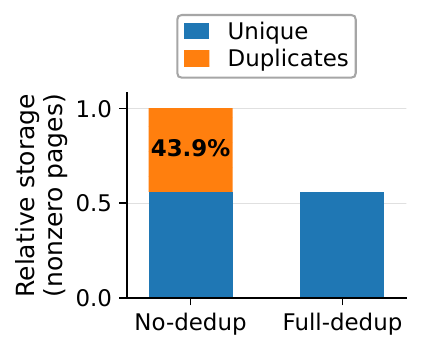}
        \caption{Nonzero snapshot storage.}
        \label{fig:back:duplication_storage}
    \end{subfigure}
    \hfill
    \begin{subfigure}[t]{0.48\linewidth}
        \centering
        \includegraphics[width=\linewidth]{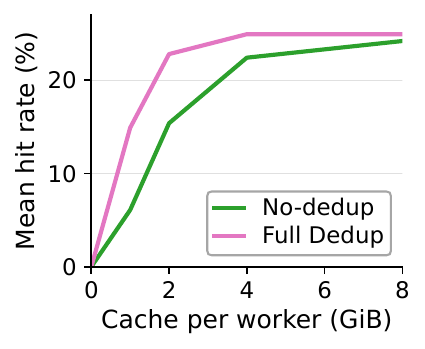}
        \caption{Working-set cache hit rate.}
        \label{fig:back:duplication_hit_rate}
    \end{subfigure}
    \caption{Cost of retaining duplicate content. (a) Duplicates occupy 43.9\% of nonzero snapshot storage without deduplication. (b) Full deduplication improves mean cache hit rate at constrained per-worker capacities; the gap narrows as capacity increases. Full deduplication is an unsafe oracle that also shares private content.}
    \Description{A stacked storage chart identifies 43.9 percent duplicates. Cache hit-rate curves compare no deduplication and full deduplication across per-worker cache capacities from zero to eight GiB.}
    \label{fig:back:duplication_cache}
\end{figure}

Duplication consumes capacity in both remote storage and worker caches. Figure~\ref{fig:back:duplication_storage} compares storage of a system with no cross-snapshot deduplication against a system with unrestricted full deduplication: duplicate copies account for 43.9\% of the nonzero page content.
% before compression.\footnote{All evaluated systems use streaming decompression upon a cold start~(\S\ref{sec:methodology}), similar to \cite{lazarev:sabre}.}
Full deduplication eliminates these copies but also shares private content; we use full deduplication as an unsafe oracle for estimating the capacity-saving opportunity, not as a deployable policy.

On a worker, duplicate copies compete with other functions' working sets for a fixed cache budget. As requests alternate among functions, admitting one working set can evict another despite their common content, causing repeated eviction and refetching. Figure~\ref{fig:back:duplication_hit_rate} quantifies this effect using the trace-driven workload described in \S\ref{sec:methodology}. A hit requires the entire working set to be local; otherwise, restoration must fetch missing content. At approximately 2~GiB per worker, full deduplication achieves a mean hit rate of about 21\%, compared with about 14\% without deduplication. The curves approach similar hit rates at larger capacities: duplication is most costly when cache space is constrained. Isolation permits reuse by the same function, but prevents its cached copy from serving equal content belonging to another tenant.

\subsection{The Measured Opportunity for Selective Sharing}
\label{sec:characterization}

% Original review comments retained for validation (before the 5999e07 edits):
% \myai{Confirm the benchmark population for these characterization plots before stating a count: the old characterization lists 13, whereas newer evaluation figures list 17.}
% We characterize 17 functions from vSwarm~\cite{vswarm}, spanning Go, Python, and Node.js; the experimental platform and workloads are described in \S\ref{sec:methodology}.
% Resolved: coauthors confirmed 17 functions in the current characterization.

% Original review comments retained for validation (before the 5999e07 edits):
% \rod{This setup is repeated, we should just mention that we use the same setup as before.}
Using the same methodology (\S\ref{sec:methodology}) as in Figure~\ref{fig:back:source_composition}, we examine recurrence within each source category by capturing page accesses using \texttt{userfaultfd}. The base snapshot is captured before adding the function image; rootfs and image references contain the corresponding filesystem content. Unmatched content is labeled \emph{Rest},\footnote{The \emph{Rest} category is distinct from the access-based division into working-set and lazy pages: unmatched pages may occur in either.} independently of whether a page belongs to the working set or the lazy remainder.
% \sout{We analyze each function's snapshot and record its working set through page faults using \texttt{userfaultfd}. The characterization matches page contents against three reference sources: a base guest-OS snapshot captured before the function image is added, the common VM root filesystem, and the function's container image. Unmatched content is categorized as \emph{Rest}. This source category is distinct from the access-based division into working-set and lazy pages: unmatched pages may occur in either. }
% 
We classify content as \emph{ubiquitous} if it appears in all tested functions, \emph{occasional} if it appears in multiple but not all functions, or \emph{unique} if it appears in only one. Source matching measures correspondence to reference bytes; recurrence measures reuse potential. Neither recurrence nor source membership alone constitutes a security approval.

\begin{figure}[t]
    \centering
    \includegraphics[width=\linewidth]{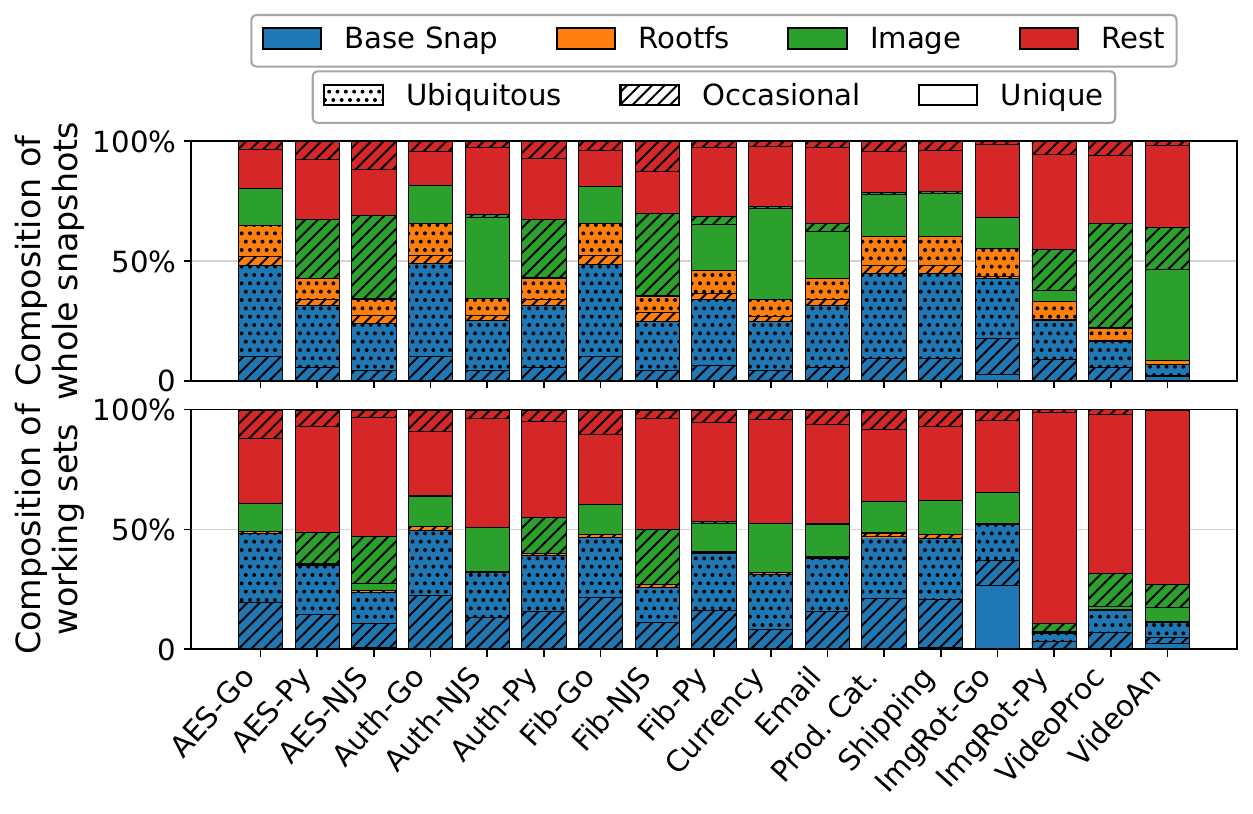}
    \caption{Existing characterization of full snapshots (top) and cold-start working sets (bottom), broken down by matching reference source and recurrence across functions. The full-snapshot breakdown excludes zero pages. Ubiquitous content occurs in all tested functions; occasional content occurs in several; unique content occurs in one. Source matches identify sharing opportunities subject to provider approval.}
    \Description{Per-function breakdowns distinguish base snapshot, root filesystem, container image, and unmatched content, as well as their recurrence across functions.}
    \label{fig:bench_breakdown}
\end{figure}

Figure~\ref{fig:bench_breakdown} expands the aggregate composition in Figure~\ref{fig:back:source_composition} into per-function source and recurrence breakdowns. The full-snapshot population describes stored content, whereas the working-set population describes content relevant to local caching and restoration. Their recurrence patterns reveal which source matches can yield reuse across functions.

% Original review comments retained for validation (before the 5999e07 edits):
% \rod{It would be good if we had some numbers to brag in the text. E.g., what is the average \% of ubiquitous and shareable data?}
Infrastructure matches account for roughly 70\% of nonzero snapshot content and 50\% of working sets (Figure~\ref{fig:back:source_composition}). Figure~\ref{fig:bench_breakdown} further shows that most base-snapshot and rootfs matches are ubiquitous, whereas image matches recur less uniformly, often among functions using the same runtime: with up to 52\% of the content that is present in all of the snapshots and 67\% of the content present in more than half of them. Similarly, the working set has up to 29\% of ubiquitous content. The aggregate source fractions quantify the opportunity; the recurrence breakdown explains why common base infrastructure is especially useful for cross-function reuse. 
% \TODO{Li, please add numbers}
% \sout{Most matching base-snapshot and rootfs pages are ubiquitous across the tested functions, so one approved copy can serve many snapshots. Container-image matches recur less uniformly, often among functions using the same language runtime. This distinction explains why providers can retain substantial sharing opportunities under a policy that excludes container-image pages. It also separates benefits from common infrastructure from any benefit that would require deduplicating arbitrary application state.}}
% \TODO{L, CHECK/fix the numbers, insert more numbers in the appropriate positions where possible/useful; please finalize this para}

\subsection{Security Model for Snapshot Sharing}
\label{sec:back:security}

Our security model focuses on confirming private snapshot content through cross-tenant deduplication. We trust the provider, its snapshot service, and its reference-approval policy. An adversarial tenant can execute arbitrary code in its own VM, influence its snapshot contents, and measure timing while co-located with a victim on the same worker. It cannot directly read victim memory or internal provider cache metadata.

An attacker can place candidate content in its snapshot and use the timing difference between local reuse and remote retrieval to infer whether a victim populated a matching cache entry. Such confirmation-of-presence attacks~\cite{suzaki:memory-disclosure,xiao:memory-dedup-security,bosman:dedup,schwarzl:remote-dedup} motivate production restrictions on cross-domain memory sharing~\cite{firecracker_docs,vmware_docs,red_hat_docs}. Encryption alone does not hide observable reuse.

% Addressed: retain the security requirement without repeating the classifier or prior-system comparison.
% \rod{All this has been said before.}
Private snapshot content must remain tenant-separated even when its bytes match; reuse by the owning function remains permitted.
% \sout{Private content must retain tenant separation in storage and caches even when its bytes are equal; reuse by the owning function remains permitted. Only complete matches to provider-approved references are eligible for sharing; unmatched or modified pages remain private. Unlike unrestricted deduplication in Medes~\cite{saxena:memory} and SnapStore~\cite{panda:snapstore}, this boundary does not permit sharing merely because two tenants hold identical content (Figure~\ref{fig:ai:duplication_comparison}).}
% 
Our model permits disclosure of the presence or use of approved content, which the operator must deem acceptable. Container-image sharing~\cite{brooker:on-demand} motivates this boundary, but does not establish protection for arbitrary guest-memory access patterns.
% \sout{Similar to existing container-based platforms that share container image content~\cite{brooker:on-demand}, our model permits detecting the presence of shared/public data.}} 
Secrets embedded in approved artifacts and secret-dependent runtime accesses are outside the scope of this work and require exclusion or additional safeguards. We do not claim protection against all shared-hardware or resource-contention timing channels; conservative approval and private treatment by default are explicit assumptions.

These observations motivate separating sharing eligibility from retrieval granularity. Approved infrastructure can be stored once and common components retained locally, while private content remains tenant-separated. Independently, working-set profiling determines which pages to retrieve eagerly. \sys carries these decisions through its storage layout, worker caches, and restoration path, with sharing eligibility governed by the provider's policy.

\section{\sys}
\label{sec:design}

% \sys's design is motivated by the snapshot characterization (\S\ref{sec:characterization}), %, which revealed two critical inefficiencies in modern serverless platforms. First, opaque chunk-based fetching causes massive network data amplification during cold starts. Second, strict isolation policies waste up to 50- 65\% of remote storage and local cache capacity by duplicating non-sensitive infrastructure components. 
%
% Original review comments retained for validation (before the 5999e07 edits):
% \myai{Align this opening with the end of Section~\ref{sec:characterization}: approved recurring infrastructure motivates shared storage and local retention; private state retains tenant identity; working-set profiling independently selects eager versus lazy retrieval. Present coalescing as an established technique, not the new contribution.}
The characterization in \S\ref{sec:characterization} motivates co-designing storage and cache layouts that enable reuse of approved infrastructure content, retain tenant identity for private content, and independently use working-set profiles to choose eager rather than lazy retrieval. \sys incorporates these decisions through three design principles:
% \sout{To deduplicate and cache non-sensitive snapshot components, \sys introduces a content-aware and cluster-wide snapshot sharing architecture. The system is built upon three key design principles:}
\begin{itemize}[leftmargin=.20cm,itemindent=5pt,labelwidth=2pt,labelsep=.2cm,align=left]
\setlength\itemsep{0pt}
    \item \textbf{Content-Match Classification over Opaque Chunks:} Rather than treating snapshot memory as an opaque blob, \sys compares each page's complete content against provider-approved sources. Only exact matches are shared; all non-matches are private.
    \item \textbf{Selective Sharing:} By decoupling the public data from the private state, \sys allows non-sensitive components to be aggressively deduplicated and pinned across worker nodes, while maintaining strict, function-exclusive isolation for sensitive components.
% Original review comments retained for validation (before the 5999e07 edits):
% \dmi{, similar to prior works~\cite{ustiugov:benchmarking,ao:faasnap}, }
    \item \textbf{Integration with Profile-Guided Fetching:} \sys incorporates working-set coalescing as in REAP~\cite{ustiugov:benchmarking} and FaaSnap~\cite{ao:faasnap} to retrieve entire working sets of pages efficiently at once, with individual 4~KB fetches on demand for the remainder. 
    % \sout{\sys uses working-set coalescing, similar to prior works, to retrieve profiled pages efficiently and individual 4~KB pages for on-demand fetching.}}
    % \leo{This retrieval mechanism complements, rather than constitutes, \sys's selective-sharing contribution.}
\end{itemize}

\begin{figure}
    \centering
    \includegraphics[width=1\linewidth]{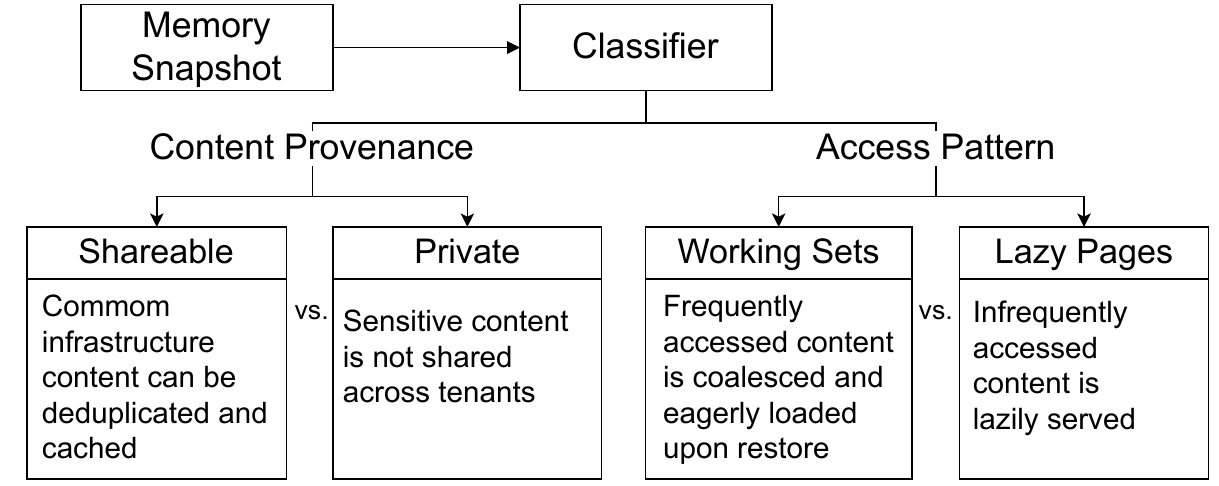}
    \caption{
    Offline snapshot classification by approved-content matching and membership in the recorded working set. A page is shareable when its current contents match an approved reference under the MD5-based match criterion; a modified page remains private if it no longer matches.
    % \sout{Offline snapshot memory classification according to exact content matches against approved sources and access frequency. A page is shared only when its complete content matches an approved source page; modified pages are private.}}
    %Content provenance detects if data is private or shareable while access patterns allow \sys to optimize retrieval by coalescing frequently accessed pages into working-sets.
    % \leo{AI-generated, replace with proper diagram}
    %based on the content of the page, it is either shared or private (to preserve security, but still get benefits of deduplication); after collecting the working set access, coalesce the pages into working set file for better retrieval performance.
    %\dmi{text needs more brief, clear phrasing. Need to specify which state category goes to which storage medium (currently, we only explain it in the text)} \TODO{essential, see my example in the comment below}
    % \dmi{Rename Pages to Lazy Pages}
    % Example:
    % * Content Dimension: Base/Rootfs/Image vs. Rest $\rightarrow$ Becomes Shared vs. Private.
    % * Access Dimension: Eager vs. On-Demand $\rightarrow$ Becomes Working Set vs. Lazy Pages.
    %
    }
    \vspace{-3pt}
    \label{fig:classifier}
\end{figure}

At a high level, \sys operates across both the control plane and the data plane. In the control plane, an offline classifier partitions the snapshot data into two dimensions (Figure~\ref{fig:classifier}): exact content matches against provider-approved sources and access patterns. A page is \textit{Shared} only if its complete content exactly matches a page from an approved source; every non-matching page is \textit{Private}. Profiling identifies pages accessed during initialization and initial requests, as in prior works~\cite{ustiugov:benchmarking,ao:faasnap}, that are coalesced into \textit{Working-Sets} that optimize retrieval; the remaining pages (\textit{Lazy}) are fetched on-demand at the page-level granularity.
% We use this profile-guided working-set coalescing as in prior work~\cite{ustiugov:benchmarking}.
In the data plane, this classification dictates the physical layout of both \textit{Remote Storage} and the worker local caches (see Figure~\ref{fig:architecture}). During a cold start, the local \textit{Controller} checks the cache for the shared components and orchestrates the targeted retrieval of the missing private components. A \textit{Page Fault Handler} then utilizes \texttt{userfaultfd} (\texttt{uffd}) to map both the cached shared state and the fetched private state into the VM's memory, allowing execution to resume once the required state is available.

\subsection{Offline Snapshot Content Classification}
\label{sec:design:classifier}
% Original review comments retained for validation (before the 5999e07 edits):
% \myai{Reconcile the implementation's hash lookup with the complete-content matching contract in Section~\ref{sec:back:security}. Explain how equality is verified and collisions handled; do not equate an MD5 match alone with a guarantee of identical contents. Describe content attribution rather than execution-history provenance.}
% Addressed: author confirmed MD5-based matching; retain this earlier question for review history.
% \myai{Implementation confirmation required: does the MD5 lookup include a byte-for-byte equality check or another collision-handling mechanism? The text below states the required complete-content contract, not an unverified claim that the prototype implements collision-safe matching.}

%The foundation of \sys is the ability to distinguish between ubiquitous infrastructure and sensitive application state. This is achieved via a control-plane component called the \textit{Classifier and Coalescer}, whose operation is depicted in Figure~\ref{fig:classifier}.

When a new function version is deployed, \sys executes an asynchronous recording phase. This profiling occurs entirely off the critical path of live user invocations, similar to the deployment-time snapshotting in AWS Lambda SnapStart~\cite{aws-snap-serv}.
% \TODO{add ref from the comment}
%source: https://www.pulumi.com/blog/aws-lambda-snapstart/ \leo{I used the one from aws guide, not pulumi}
To execute the recording, the control plane provisions a dedicated profiling MicroVM on a worker node. This instance is bootstrapped by restoring a VM from a generic, non-sensitive base image (comprising the booted guest OS and rootfs), then takes a snapshot after mounting the function's specific container image. To precisely capture the memory access patterns, a local profiling agent monitors the VM's guest memory using \texttt{userfaultfd} (\texttt{uffd}). As the VM initializes the language runtime and loads the application code, every access to an unmapped memory page triggers a page fault, which the agent intercepts and logs.

To record the working set for the profiling requests, the platform then routes a series of initial requests, such as developer-provided test payloads or synthetic health probes, to the running function~\cite{ao:faasnap,ustiugov:benchmarking}
% \dmi{L, add ref}
The agent continues to trace all \texttt{uffd} faults during this request-serving window. The aggregate set of all pages faulted in during both the initialization and the initial execution phases is explicitly recorded; these observed pages define the function's profiled \emph{Working Set}. Later requests may access additional pages, which are served on demand. 
% \sout{this precise collection of pages defines function's \emph{Working Set}.}}

Once recorded, the offline classifier computes an MD5 digest over each non-empty snapshot page and compares it with the digests of provider-approved reference pages, including the base snapshot and rootfs. The implementation uses equal MD5 digests as its page-content match criterion.
% \sout{Once recorded, the offline Classifier analyzes every non-empty page in the snapshot, comparing its MD5 hash against a registry of pages from known, provider-approved sources, specifically the base snapshot and the rootfs.}} 
Providers may also approve the exact contents of selected libraries or container images. The classification tests the snapshot page's current contents, not which image originally supplied it:
% \sout{This is a complete-content comparison, not provenance tracking:}} 
only a page whose full content matches an approved source page is tagged \emph{Shared}; every unmatched page is tagged \emph{Private}, including the pages initially supplied by a container image whose modified contents no longer match approved references. 
% \sout{including a page initially supplied by a container image but modified during initialization or execution.}} 
Second, pages accessed during the recording phase are tagged as the \emph{Working Set}, while the unaccessed remainder is tagged as \emph{Lazy}.

\subsection{Memory Management Granularity}
\label{sec:design:granularity}
% \myai{Integration check: main.tex no longer includes chunk-size-ai.tex, but Evaluation still references sec:char:chunks for its 128 KB baseline. Restore the relocated study or revise that reference when finalizing Evaluation.}
Current production systems segment memory into rigid chunks (e.g., 512~KB), leading to data amplification and retrieval-granularity mismatch~\cite{aws_compute_optimizing_lambda}. To avoid this established limitation, \sys incorporates profile-guided working-set coalescing~\cite{ustiugov:benchmarking} alongside its content-match classification.

\textbf{\textit{Working Sets for Eager Loading:}} Following prior profile-guided retrieval designs~\cite{ustiugov:benchmarking}, the recorded working-set pages are coalesced into files that preserve their shared or private classification.
% \sout{the pages identified as the Working Set are coalesced into a contiguous file.} 
Coalesced working-set transfers amortize retrieval overhead without fetching surrounding pages merely because they share a chunk. The profile predicts which pages to load eagerly; accesses outside that profile use the lazy path. 
% \sout{By retrieving the working set as a single unit rather than fragmented chunks, \sys maximizes network bandwidth utilization during the critical path of a cold-start, transferring only the exact bytes needed to resume execution without the bloat of surrounding unaccessed memory.}
% \leo{This retrieval mechanism complements \sys's selective-sharing contribution.} %\cz{not clear the method to achieve the target of transferring only exact bytes needed.}

\textbf{\textit{Lazy Pages for On-Demand Fetching:}} The pages that are not accessed during the recording phase are not coalesced. Instead, they are stored and managed at the granularity of individual 4~KB pages. Because these pages fall outside the predicted working set, they are fetched on demand if a post-restore page fault occurs, limiting transfer size without requiring the profile to cover every subsequent access.
% \sout{minimizing network amplification and preventing bandwidth contention.}}
% \TODO{add accuracy of ws prediction. cite reap}

% Original review comments retained for validation (before the 5999e07 edits):
% \myai{Make the four shared/private and working-set/lazy classes explicit in storage and cache policy. Distinguish approved content from content actually pinned; qualify small pinned-footprint claims with their measured scope. Describe cache misses for shared components as well as private ones, and avoid claiming restoration is instantaneous.}
\subsection{Co-designed Cache and Storage Layout}
\label{sec:design:codesign}

\begin{figure}
    \centering
    \includegraphics[width=\linewidth]{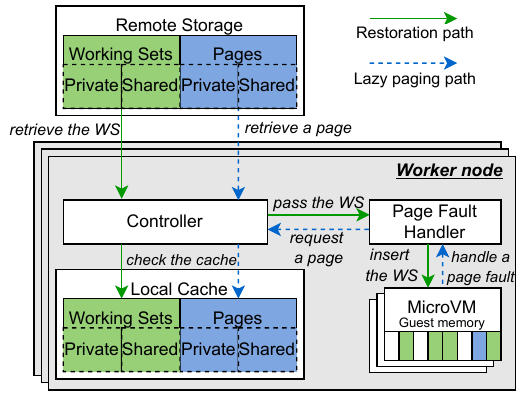}
    \caption{
    \sys's storage and cache co-design that combines selective sharing and dual-granularity fetching.
    % \dmi{replace the storage bucket name "Pages" with "Lazy Pages". (e.g., Lazy Pages Storage, Shared Lazy Pages, Private Lazy Pages).}
    % \rod{ Shared Working Sets, Private Working Sets, Shared Pages, and Private Pages should appear similar to how the local cache has it.}
    % \leo{1-4 are for restoration (retrieves working set if needed and inserts); a-d are for page faults (initiated from the vm, goes to the controller that retrieves the page or serves from cache)}
    }
    \vspace{-10pt}
    \label{fig:architecture}
\end{figure}

To leverage the classifier output and effectively apply selective sharing and dual-granularity fetching, \sys offers a co-designed storage and caching hierarchy (see Figure~\ref{fig:architecture}).

Sharing eligibility is determined from snapshot-page contents before compression. 
% \sys and all evaluated baselines use streaming compression for snapshot retrieval (\S\ref{sec:methodology}). 
On a cold start, the worker node overlaps transfer with decompression, so it must extract memory contents before loading them into guest memory. Compression changes the storage and transfer representation, not a page's shared/private ownership or working-set/lazy classification.

\subsubsection{Remote Storage}
% Original review comments retained for validation (before the 5999e07 edits):
% \TODO{reduce usage of md5 - it is implementation detail, design should be general hash-match}
The centralized repository is split into \textit{Working Set Storage}, holding the coalesced files for the restoration path, and \textit{Lazy Pages Storage}, holding individual pages for lazy retrieval. Shared lazy pages use content-based identifiers to enable reuse across snapshots; identifying equal content must satisfy the classifier's content matching contract.
% \sout{To maximize storage efficiency while enforcing isolation, remote storage is content-addressed. Lazy Pages are indexed by their MD5 hashes, allowing the storage service to perfectly deduplicate shared components globally across the entire cluster.}} 
Private components, however, are indexed alongside their unique tenant and function version identifiers to guarantee tenant-level isolation. In sum, the remote storage layout is logically divided into four distinct sections: Shared Working Sets, Private Working Sets, Shared Lazy Pages, and Private Lazy Pages.

\subsubsection{Worker Local Caches} 
On each worker node, the local cache mirrors the remote structure but is logically partitioned based on content provenance. Crucially, \sys maps guest memory via \texttt{userfaultfd} (\texttt{uffd}), hence the local cache operates entirely in user space. This architectural flexibility allows the local agent to implement a custom, provenance-aware caching/eviction policy:
% that explicitly protects ubiquitous infrastructure:

\textbf{\textit{Shared partition:}} stores content approved for cross-tenant reuse. The controller pins common base-snapshot and rootfs components to retain their reuse across functions. In our measurements, this pinned base state occupies less than 1\% of cache capacity; this is a workload-specific footprint, not a bound on an arbitrary approved registry.
% \sout{\textbf{\textit{Shared partition:}} stores shared pages originating from non-sensitive snapshot components that are deemed safe to deduplicate; they are explicitly pinned in the local cache and shared freely across functions and tenants. The cardinality of these base infrastructure components is extremely low; thus pinning them consumes a small, bounded amount of memory while guaranteeing immediate availability (less than 1\% of the cache content in our experiments).}}

\textbf{\textit{Private partition:}} stores every page that does not exactly match an approved source page, including sensitive, application-specific pages, and unapproved libraries. These pages remain strictly isolated per function. Private working sets are managed with LRU eviction under cache pressure, while pinned shared components remain resident. Sharing approval determines whether content may be reused across tenants; pinning separately determines which approved components are retained.
% \sout{Under high memory pressure, the local cache applies a standard LRU policy exclusively over this partition, evicting these single-tenant private pages while the shared partition remains pinned.}}

% Follow-up retained: the manuscript does not specify eviction of approved image working sets.
% \myai{Author confirmation: are approved container-image working sets pinned or evictable? Specify their policy without conflating all Shared content with the pinned base/rootfs components.}
% \leo{the answer is: in evaluation, we pin. but overall, can be evictable. with only a few actual images we either way wouldn't see the difference. }

\subsubsection{Restoration Flow}
During a cold start, the controller checks which shared and private working-set components are resident and retrieves the missing components from remote storage. The page-fault handler restores the working set from these cached files. Subsequent accesses outside the recorded working set trigger individual page fetches through the local cache, with remote retrieval on a miss. Shared content remains eligible for cross-tenant reuse; private content retains its tenant identity throughout this path. 

\subsection{Enforcing the Sharing Boundary}
\label{sec:design:security}

Our security model (\S\ref{sec:back:security}) defines the trusted provider, the attacker's capabilities, and the permitted leakage. \sys enforces that boundary through hash-based matching against approved references and tenant-scoped identifiers for private content. Equal private pages are not reusable across tenants; approved pages may be shared per provider policy. 
% \sout{Private components are never deduplicated across tenants, even when their contents are equal. This prevents their reuse from directly creating a cross-tenant deduplication hit; it is not a guarantee against every timing effect of shared resources.}

Source approval does not establish that an artifact contains no secrets or that its run-time accesses are secret-independent. Components with these risks require exclusion or additional safeguards, consistent with \S\ref{sec:back:security}. Screening tools~\cite{yuan:cacheql,iyer:cfar,geimer:sidechannel-tools}
% \TODO{L, can you check if these refs are correct? I suspect hallucinations here, not sure}
may inform that policy, but are not protections established by the page classifier itself.
% \sout{A separate concern is the secret-dependent access pattern in common infrastructure. \sys is not enough on its own to resolve this issue, but can be used in conjunction with additional screening during the deployment phase for shareable libraries that may exhibit a secret-dependent pattern, and by excluding individual pages from sharing to avoid this type of side-channel.}}

We emphasize that \sys is intended for use by operators as a framework for defining dynamic policies that determine which components are safe to share. 
%
%\textbf{\textit{Dynamic Security Models:}} 
%\sys's design empowers cloud providers to dynamically adjust the ``sensitive'' boundary. 
% \TODO{clearer on the revocation process, point to results of no-image-sharing}
% Original review comments retained for validation (before the 5999e07 edits):
% \myai{For the revocation discussion below, separate implemented behavior from proposed policy mechanisms. Specify how existing shared cache entries and snapshot references are handled before asserting instant revocation or restored safety; retain the conservative-policy experiment as evidence of cost, not of security.}
Operators can revise the registry of approved reference content as their security assessment changes. If a newly disclosed vulnerability makes a component's presence sensitive, operators must withdraw its sharing eligibility. Enforcing that policy requires handling existing cached copies and snapshot references as well as future classifications. 
% \sout{Operators can leverage page origin and access patterns to safely expand the shared registry with popular, non-sensitive language libraries (e.g., standard Python packages), boosting local cache hit ratios. On the other hand, if a zero-day vulnerability is discovered, making the mere confirmation of a library's presence a security risk for a tenant, the provider can instantly revoke its safe status.}}
% The classifier seamlessly downgrades those pages to the private partition, restoring strict structural isolation without requiring systemic architectural changes.
% Original review comments retained for validation (before the 5999e07 edits):
% \dmi{I am a bit confused: we evaluate both with image and w/o, right? also, why the ref is to the entire methodology? I think it can be confusing}
A possible revocation procedure is to disable cross-tenant cache reuse for the affected content while regenerating snapshot metadata to point to tenant-private copies.
% This describes the required policy transition, not a measured instantaneous-revocation guarantee.
This procedure should close the possible timing channel attack nearly instantaneously.
We evaluate both image-sharing policies: \sys includes deduplication of container-image content, whereas \sys- excludes it, as the lower and upper bounds of possible policies.
% Their cache savings relative to the Sabre baseline are 38.4\% and 31.2\%, respectively
The most restrictive image-sharing policy leads to only a 10\% increase in cache size
(\S\ref{sec:eval:storage}); the tail-slowdown comparison appears in \S\ref{sec:eval:e2e}. These comparisons quantify the performance cost of a stricter policy, not revocation correctness or latency.

\subsection{Implementation}
\label{sec:design:implementation}

We implement \sys atop vHive~\cite{ustiugov:benchmarking}, extending its Knative and Kubernetes infrastructure. Function sandboxes execute as Firecracker MicroVMs~\cite{agache:firecracker}. We bypass Knative's standard container initialization, redirecting cold-starts directly into our custom snapshot retrieval pipeline.

\textbf{\textit{Controller:}}
We modify Knative's Activator to intercept cold-start requests and route them to our custom node-level controller. Written in $\sim$3,000 lines of Go, the controller orchestrates the control-plane restoration logic and executes the offline Classifier and the coalescer routine. The controller has no state that cannot be recovered from the global cluster state, making it easy to reconstruct after failures.

\textbf{\textit{Remote Storage:}}
We build the centralized repository using MinIO. To optimize the high-throughput, small-I/O workload required for serving individual 4~KB pages, we format the underlying storage nodes with the XFS filesystem, which empirically outperformed standard \texttt{ext4} in our small-object retrieval evaluations.
% \kev{I have a really small concern about this because it looks like XFS is a big part of the better results of \sys, "Chunks" baseline will inevitably benefit less than \sys on small-object oriented file systems like XFS}.
We set up remote storage to provide ms-scale chunk retrieval latencies, comparable to datacenter/AZ-level storage tier used for snapshots in AWS Lambda~\cite{brooker:on-demand}.
We apply compression to all of the content in the remote storage to match the storage savings and network transfer performance to match state-of-the-art works~\cite{lazarev:sabre}.
% , which serves 99.9\% of image chunks.
% \TODO{Refer to the AWS doc that talks about single-digit ms retrieval time in their depl, similar to ours. In general, weave in the argument that our remote-local hierachy is similar to AWS.}

\textbf{\textit{Local Cache:}} 
The node-level cache runs as a user-space daemon managing NVMe drives. As designed in \S\ref{sec:design:codesign}, the daemon explicitly pins shared components. Production environments exhibit extremely low cardinality for the base state,\footnote{Major providers typically support only 5-7 standard language runtimes~\cite{aws_lambda_containers,azure_functions_containers,gcr_deploying}. Developers overwhelmingly prefer these official images for their optimized performance and regular vulnerability checks~\cite{azure_functions_containers,gcr_buildpacks_vs_dockerfile,pcg_lambda_containers,aws_compute_optimizing_lambda}.} thus this pinned shared state consumes a negligible fraction ($<1\%$) of the node's cache capacity in our evaluations. The daemon enforces strict LRU eviction for all remaining private and on-demand state.

\textbf{\textit{Page Fault Handler:}}
% \dmi{ai simplied the description here, CHECK}
Implemented in $\sim$1,000 lines of Go, the handler interfaces with Firecracker's \texttt{userfaultfd} API. During eager restoration, it supplies decompressed working-set page contents to guest memory. For lazy accesses, it resolves faulting guest addresses through the recipe metadata and obtains the required page through the cache or remote retrieval path.

\section{Evaluation}
\label{sec:eval}

We evaluate \sys to demonstrate that selective, structural snapshot deduplication enables rapid serverless cold-starts without compromising tenant isolation. 
%Reflecting the hierarchical structure of our experiments, 
Our experiments focus on answering thee main questions. First, what is the end-to-end tail latency and local cache hit efficiency offered by \sys (\S\ref{sec:eval:e2e})? Second, what is \sys's storage impact and performance breakdown during a cold-start (\S\ref{sec:eval:per-comp})? Third, validate working set estimation and measure how sensitive is \sys to network bandwidth (\S\ref{sec:eval:sensitivity})?

% Reordered methodology proposal: retain all existing review markup.
% Sequence: setup, workload, common compression, system variants, baselines.
\subsection{Methodology}
\label{sec:methodology}

\textbf{\textit{Hardware and Software Setup:}}
% \leo{we have two setups, one with single worker + single minio node, another with 10 nodes total. latter is only used in 7.5, former for everything else}
We evaluate \sys on a CloudLab~\cite{cloudlab} cluster comprising 10 \texttt{c6620} bare-metal nodes, except for latency and footprint profiling experiments, which are performed on a single worker and a single MinIO node. Each node is equipped with a 28-core Intel Xeon Gold 5512U (Emerald Rapids) processor clocked at 2.1~GHz, 128~GB of DDR5 RAM, and a 780~GB NVMe SSD. To accurately mimic a multi-tenant production cloud environment—where network bandwidth must be strictly provisioned to guarantee performance for user logic—we throttle the 100~Gbps NICs to 10\% (10~Gbps) for all snapshot state downloads, unless otherwise specified.
% We evaluate \sys on an AWS EC2 cluster comprising 10 \texttt{c7i.metal-24xl} bare-metal nodes, except for latency and footprint profiling experiments, which are performed on a single worker and a single MinIO node. Each node is equipped with a 48-core Intel Xeon Platinum 8488C processor clocked at 2.1~GHz with SMT disabled, 192~GB of DDR5 RAM, a 512~GB NVMe SSD, and 4 IAA accelerators. To accurately mimic a multi-tenant production cloud environmentm where network bandwidth must be strictly provisioned to guarantee performance for user logic, we throttle the 25~Gbps NICs to 10\% (10~Gbps) for all snapshot state downloads, unless otherwise specified.
For the compression comparison in~\S\ref{sec:eval:compression}, we use a phoenixNAP~\cite{phoenixnap} \texttt{s5.x6.c3.medium} bare-metal instance equipped with a 24-core Intel Xeon 6527P processor clocked at 2.1 GHz with SMT disabled, four Intel IAA devices, 256~GB of DDR5 memory, and two 2~TB NVMe SSDs.
% \rod{If this is used in single experiment, mention which section.}

% \leo{new nodes: c7i.metal-24xl, 96vCPU, 192GB RAM; 512 GiB of EBS attached as storage}
% lik{}

Worker nodes run Ubuntu Linux~24.04 (kernel v6.8). The cluster infrastructure is orchestrated using vHive~v1.8.1, with Knative~v1.13 running on Kubernetes~v1.29. Function sandboxes are instantiated as Firecracker MicroVMs~v1.13 with kernel~v5.10 inside, each configured with 1~vCPU and 512~MB of RAM (except for video processing and video analytics, which need 2 and 3~GiB, respectively).
% \footnote{Due to the container image sizes, we increase the allocations for the two image-rotation functions and the video-processing function to 2~GiB each, while the video-analytics function uses 3~GiB.} All MicroVMs run Debian~12 with Linux kernel~v5.10.
% , running Debian~12 Linux guest with kernel~v5.10.

\textbf{\textit{Workload:}}
% \sout{To provide representative application logic, we map} 
% \dmi{brutally shortened to the essence, CHECK}
For the trace-driven experiments (\S\ref{sec:eval:e2e}), we sample 700 Azure Functions~\cite{shahrad:serverless} invocation profiles using In-Vitro~\cite{ustiugov:invitro}, preserving the trace's inter-arrival-time and execution-duration distributions. We map 17 vSwarm~\cite{vswarm} benchmarks written in Go, Python, and Node.js to these profiles by execution duration and memory footprint, replicating them as independent functions. We replay the resulting workload with In-Vitro~v1.1~\cite{invitro:github} to exercise cache contention and concurrent snapshot retrieval. For other experiments, we measure each function's snapshot footprint and cold-start latency individually at one request per second. 

\textbf{\textit{Common Compression Settings:}}
We apply compression to all evaluated systems as state-of-the-art methodology of improving the fetching~\cite{lazarev:sabre}.
% \leo{We use IAA-accelerated decompression overlapped with network transmission latency.}
We use \emph{zstd} as fast, streaming software compression, overlapping with network transmission latency. We use 8 decoders to parallelize the decompression. We evaluate the impact of choosing this specific compression against IAA-accelerated compression in~\S\ref{sec:eval:compression}.
% \dmi{subject to change}

\textbf{\textit{\sys Configurations:}}
We evaluate two versions of \sys: \emph{\sys}, which deduplicates all the common components, including container image,
% \rod{This sentence seems to imply that \sys only deduplicates container-image pages. I don't think this is the case. We should probably present \sys as the system described above and a stricter version of \sys that does not deduplicate container-image pages.}
% \dmi{be careful here: we are not a provider to approve things; be specific. you can call conservatively assumes and optimistically assumes}
and \emph{\sys-}, which does exclude the container-image content. These configurations estimate the best- and worst-case performance after applying the mitigations discussed in~\S\ref{sec:design:security}.

\textbf{\textit{Baselines:}}
We compare \sys with four baseline storage management and retrieval strategies for guest-memory snapshots. Chunks, Pages, and Sabre retain separate content across functions and permit reuse by requests of the same function; Full-Dedup is the explicitly unsafe exception. \sys and \sys- share only provider-approved content, according to the policies above.
% \rod{Edited the paragraph above.}
% The original sharing-rule paragraph is preserved unchanged in the previous version.

\begin{itemize}[leftmargin=.20cm,itemindent=5pt,labelwidth=2pt,labelsep=.2cm,align=left]
\setlength\itemsep{2pt}
    \item \emph{Chunks:} A chunk-based transfer unit and storage layout in both the remote storage and local cache, mirroring the design of production systems like AWS Lambda SnapStart~\cite{aws-snapstart-optimization}. We configure the chunk size to 128~KB.\footnote{We empirically determined 128~KB chunks to be the best-performing size because they saturate I/O bandwidth while limiting unnecessary data movement due to the chunk layout.}
    % (as characterized in \S\ref{sec:char:chunks}) \TODO{CRIT}\dmi{We removed this study, add a short (foot)note about why 512}.
    \item \emph{Pages:} A page-based transfer unit and storage layout. This baseline minimizes redundant state transfers by fetching state at the strict granularity of individual 4~KB pages, perfectly matching the underlying Linux memory management system.
    \item \emph{Sabre:} A system representing state-of-the-art serverless cold start optimization system~\cite{lazarev:sabre}. Implemented by integrating the working set into a single file~\cite{ustiugov:benchmarking,ao:faasnap} and
    applying streaming zstd compression. We evaluate the choice of the compression against original IAA-accelerated compression separately in~\S\ref{sec:eval:compression}.
    % applying IAA-accelerated compression.
    On a cache miss, it retrieves the full working set file and decompresses it on the fly. Remaining on-demand page faults are served as individual 4~KB granularity.
    % \dmi{keep in mind, if we call it Sabre, we have to start with the paper's design/setup, then you can add REAP or etc. this can be a sensitive issue} A working-set-based transfer unit and layout, representing REAP-style restores~\cite{ustiugov:benchmarking,ao:faasnap}. The predicted working set is eagerly fetched as a coalesced file, while the remaining individual pages are served strictly on-demand. Working sets are captured after five processed invocations for each function. We apply streaming \emph{zstd} compression, overlapped with fetching, to represent the more recent Sabre design~\cite{lazarev:sabre}.
    % \dmi{explain why you took zstd; subject to updates}
    \item \emph{Full-Dedup:} A system that employs full deduplication, representing prior works~\cite{saxena:memory,panda:snapstore} and estimating best-case performance. By construction, it does \emph{not} satisfy our security model (\S\ref{sec:back:security}), acting as an upper bound for performance.
\end{itemize}

\subsection{End-to-End Performance}
\label{sec:eval:e2e}
% \leo{all eval is updated to match new figures, need to fill in the gaps}

\begin{figure}
    \centering

    \begin{subfigure}[t]{0.49\linewidth}
        \centering
        \includegraphics[width=\linewidth]{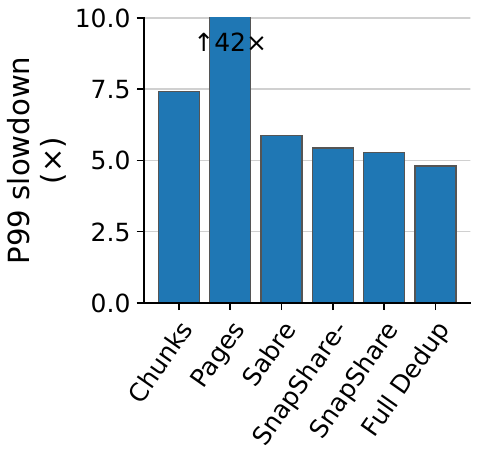}
        \caption{End-to-End slowdown.}
        \label{fig:eval:e2e_slowdown}
    \end{subfigure}
    \hfill
    \begin{subfigure}[t]{0.49\linewidth}
        \centering
        \includegraphics[width=\linewidth]{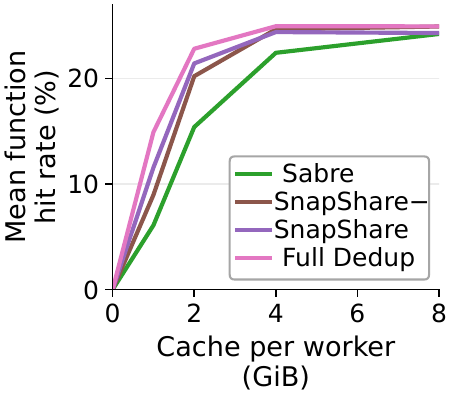}
        \caption{Working-set cache hit rate.}
        \label{fig:eval:snapshot_hit_rate}
    \end{subfigure}

    \caption{
        (a) Aggregate 99\textsuperscript{-th} percentile request slowdown on the 700-function trace under a 4~GB per-node cache limit. \sys reduces tail latency by combining dense caching with optimized remote fetch payloads. Lower is better.
        (b) Cache hit rate under varying per-node capacity limits for a realistic 700-function trace, spanning increasing levels of deduplication from \emph{Sabre} through \sys-, \sys, and unsafe \emph{Full Dedup}. Deduplication improves cache density under constrained capacity, while all configurations approach a similar saturation point. Higher is better.
        % \leo{maybe combine with the cache hit rates as subfigures}
    }
    \label{fig:eval:cache_performance}
\end{figure}

% To evaluate end-to-end performance under realistic cluster dynamics, we execute a production-derived trace of 700 serverless functions across a cluster of 8 worker nodes.
We first quantify the bottom-line reduction in end-to-end tail slowdown for user requests. To explain these performance gains, we then examine how our structural deduplication improves local cache hit rates by maximizing storage density.

\subsubsection{Tail Request Slowdown}
To evaluate the ultimate impact of \sys's architectural optimizations on user-facing performance, we measure the worst-case slowdown across all evaluated systems under a strictly fixed cache limit of 4~GB per worker node. Figure~\ref{fig:eval:e2e_slowdown} illustrates the aggregate 99\textsuperscript{-th} percentile slowdown.\footnote{We calculate the per-invocation slowdown as the ratio of the end-to-end delay to the unloaded ideal execution latency for the function. We then compute the 99\textsuperscript{-th} percentile across all invocations for each function, taking the geometric mean of these tail slowdowns across the entire trace.}
\sys reduces tail slowdown to 5.2, a 10\% reduction relative to \emph{Sabre} and a 29\% reduction relative to \emph{Chunks}. It performs close to the more restrictive \sys- configuration and the unsafe \emph{Full Dedup} baseline: within 8\%. \sys uses the limited local cache capacity to maximize immediate hits while minimizing the payload fetched on a miss, a network dynamic explored in \S\ref{sec:eval:latencies}. By bounding restoration penalties on both paths, \sys improves tail latency in multi-tenant serverless deployments.

\subsubsection{Cache Hit Rate Efficiency} 
The end-to-end latency improvements observed above are heavily driven by cache density under strict capacity limits. To demonstrate this, we vary the local cache capacity for configurations spanning the deduplication spectrum: \emph{Sabre}, \sys-, \sys, and unsafe \emph{Full Dedup}. We omit \emph{Chunks} and \emph{Pages} because their per-chunk and per-page tracking mechanisms produce incomparable hit-rate definitions.
In this experiment, a cache hit occurs if the restoration procedure finds the entire working set contents (both shareable and private) in the local worker cache. On the other hand, a cache miss happens if part of the working set needs to be fetched from remote snapshot storage.
% Figure~\ref{fig:eval:snapshot_hit_rate} demonstrates that \sys consistently achieves a higher cache hit rate under identical capacity constraints. 
Figure~\ref{fig:eval:snapshot_hit_rate} shows that increasingly aggressive deduplication improves the hit rate when cache capacity is constrained: \sys- shares provider-approved base infrastructure, \sys additionally shares approved container-image pages, and \emph{Full Dedup} provides the unsafe upper bound. These configurations primarily differ before saturation. Once the cache is large enough to retain the workload's reusable working sets, all deduplicating configurations converge to a similar saturation point, with a maximum mean function hit rate of 24\%. Thus, deduplication changes how efficiently the cache reaches the workload-imposed ceiling, rather than materially changing that ceiling.
The saturation comes for all three deduplication policies at cache sizes within 10\% of each other, and at half the cache size of \emph{Sabre}.

\subsection{Per-Component Analysis}
\label{sec:eval:per-comp}

% \TODO{add lead-in}

To pinpoint the exact sources of \sys's improvements, we analyze both storage requirements and restoration performance. We start by estimating the storage and cache footprints. Following this, we decompose the cold-start latency into stages and evaluate which ones our system accelerates, supported by an analysis of the network retrieval sizes.

\subsubsection{Storage Requirements}
\label{sec:eval:storage}

We first analyze the static storage footprint of the collected snapshots and coalesced working sets to quantify the space savings enabled by our selective deduplication architecture.

\begin{figure}
    \centering

    \begin{subfigure}[t]{0.49\linewidth}
        \centering
        \includegraphics[width=\linewidth]{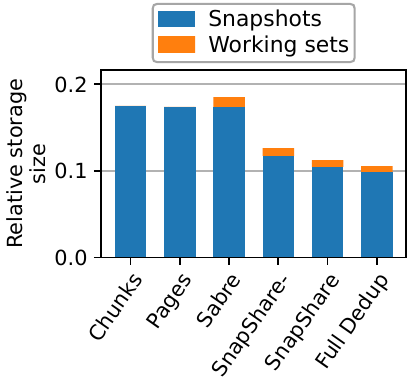}
        \vspace{-15pt}
        \caption{Remote storage usage}
        \label{fig:eval:snap_sizes_optimizations}
    \end{subfigure}
    \hfill
    \begin{subfigure}[t]{0.49\linewidth}
        \centering
        \includegraphics[width=\linewidth]{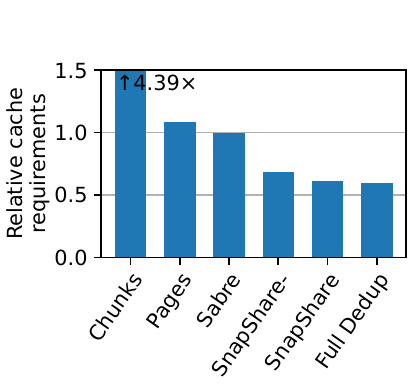}
        \vspace{-15pt}
        \caption{Local cache requirements}
        \label{fig:eval:ws_sizes_optimizations}
    \end{subfigure}

    \caption{
        % Storage requirements for the snapshots and working set files of 13 real benchmarks. Values normalized to storing the full unmodified snapshot files.
        % \TODO{add insecure baseline to this fig and to the text}
        % \leo{combine with fig11 (cache sizes) as subfigures}
        % \lik{(a) Remote storage capacity requirements for full snapshots and coalesced working set files across 17 representative serverless benchmarks. Values are normalized against the storage footprint of full snapshot files. Lower is better. (b) Worker-node local cache capacity required to store the active working sets across the benchmark suite. Values are normalized against the \emph{Sabre} baseline footprint. By explicitly pinning only low-cardinality shared components, \sys approaches the footprint of full deduplication. Lower is better.}
        (a) Remote storage capacity and (b) worker-node local cache capacity requirements for storing snapshots and coalesced working set files across 17 representative serverless benchmarks. Values are normalized against the full snapshot sizes and \emph{Sabre} baseline footprint for (a) and (b) respectively. By deduplication and explicitly pinning only low-cardinality shared components, \sys approaches the footprint of full deduplication. Lower is better.
        % \dmi{the label on the left bar should be 5.1$\times$}
        % \TODO{add the full dedup here as well}
    }
    \vspace{-7pt}
    \label{fig:eval:sizes_optimizations}
\end{figure}

\textbf{\textit{Remote Storage:}}
Figure~\ref{fig:eval:snap_sizes_optimizations} illustrates remote-storage requirements normalized to full snapshot size. \emph{Chunks}, \emph{Pages}, and \emph{Sabre} require 17.5\%, 17.4\%, and 18.5\% of the normalized capacity, respectively. Avoiding container-image deduplication increases \sys's footprint from 11.2\% to 12.6\% for \sys-.
% \dmi{what are the clean pages? we already deleted zeroed ones, right?} \leo{not by default, values normalized to full snapshot, including all clear pages untouched. page- and chunk-based get their benefits from not storing those (but chunks require full chunk empty, not just a page inside)} intra-snapshot. 
The coalesced working-set component adds 6.2\% to \emph{Sabre}'s snapshot storage. By globally deduplicating non-sensitive shared pages across all 17 benchmarks, \sys reduces storage by 39\% relative to \emph{Sabre} and remains within 6\% of the unsafe \emph{Full Dedup} lower bound.

\textbf{\textit{Local Cache Footprint:}} Figure~\ref{fig:eval:ws_sizes_optimizations} shows the capacity required to cache active working sets. \emph{Chunks} requires 4.39$\times$ the \emph{Sabre} footprint because of its granularity mismatch, while \emph{Pages} requires 1.1$\times$. By pinning shared components once per node, \sys reduces the cache footprint by 38.4\% relative to \emph{Sabre}; \sys- achieves a 31.2\% reduction. \sys remains within 3.2\% of unsafe \emph{Full Dedup}.

\subsubsection{Cold-Start Latency}
\label{sec:eval:latencies}

\begin{figure}
    \centering
    \includegraphics[width=\linewidth]{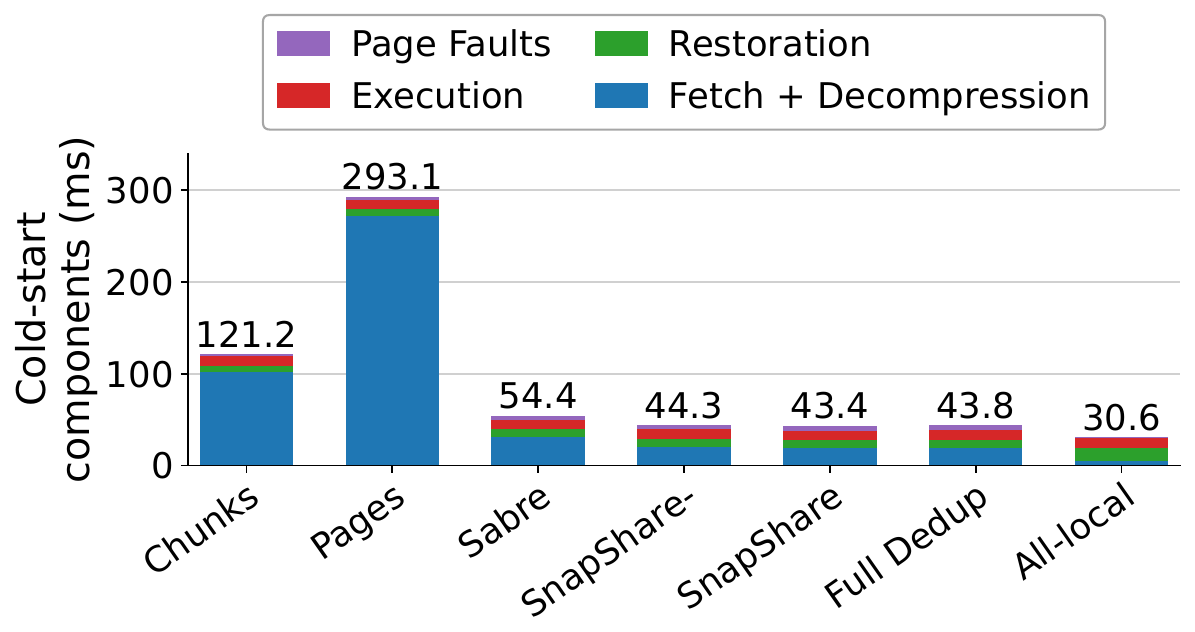}
    \vspace{-25pt}
    \caption{
    % Breakdown of the cold start delays in systems for synthetic functions in worst cases (fetching data from remote). Note that the benefit comes from shrinking fetch latency.
    Cold-start latency breakdown when restoring the \texttt{aes-go} benchmark entirely from remote storage. Every configuration uses streaming compression, with decompression overlapped with the network fetch and reported as one combined stage. \sys reduces this stage primarily by minimizing the compressed data transferred.
    % \TODO{this chart should be augmented with the all-local measurements, i don't think we need this anywhere else}
    % \TODO{Add total labels above the bars}
    % \leo{increase font size a bit}
    \vspace{-5pt}
    }
    \label{fig:eval:remote_delay_breakdown}
\end{figure}

\begin{figure}
    \centering
    \includegraphics[width=\linewidth]{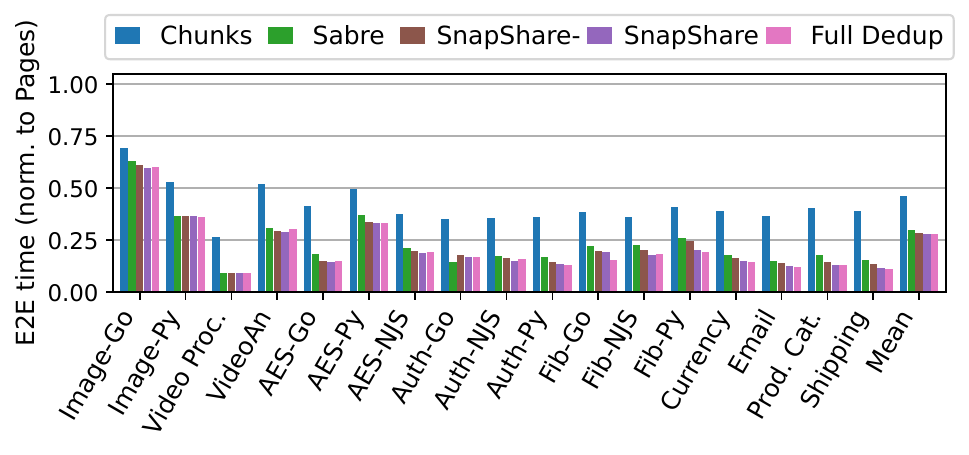}
    \vspace{-20pt}
    \caption{
    End-to-end cold-start latency across 17 diverse serverless benchmarks in a worst-case remote-only fetch scenario. Values are normalized to \emph{Pages} (not shown). \sys outperforms \emph{Chunks} and \emph{Sabre} by securely eliminating redundant infrastructure downloads. Lower is better.
    % \kev{SplitShare on the fig}
    % \dmi{remove Pages bars}
    % \TODO{normalize to Pages and do not show them}
    % \leo{increase font size}
    }
    \vspace{-5pt}
    \label{fig:eval:e2e_time}
\end{figure}
To evaluate worst-case cold-start performance, we measure restoration latency while fetching required state from remote storage and bypassing the local cache. We apply streaming compression to every evaluated configuration. Decompression proceeds concurrently with the network transfer, so the breakdown reports their overlapping critical-path cost as a single \textit{Fetch + Decompression} stage rather than adding two serial costs. The remaining stages are \textit{Restoration}, which instantiates the VMM and configures \texttt{uffd} mappings; \textit{Execution}, the benchmark's ideal execution time; and \textit{Page Faults}, the cost of serving on-demand remote faults.

Figure~\ref{fig:eval:remote_delay_breakdown} shows the cold-start breakdown for the representative string-encryption benchmark \texttt{aes-go}. Total latency is 121.2~ms for \emph{Chunks}, 293.1~ms for \emph{Pages}, and 54.4~ms for \emph{Sabre}. Because all configurations use the same overlapped compression pipeline, the differences in \textit{Fetch + Decompression} primarily reflect the amount and layout of compressed state transferred rather than whether compression is enabled. Removing redundant infrastructure transfers lowers total latency to 44.3~ms for \sys- and 43.4~ms for \sys, close to the unsafe \emph{Full Dedup} result of 43.8~ms. The 30.6~ms \emph{All-local} result shows the remaining restoration, execution, and page-fault cost when no remote fetch is required.

Figure~\ref{fig:eval:e2e_time} extends the analysis across all benchmarks. \sys reduces end-to-end cold-start time by 39.8\% relative to \emph{Chunks} and by 27.5\%, with a mean reduction of 6.1\%, relative to \emph{Sabre}. The gap between \sys- and \sys quantifies the additional benefit of deduplicating approved container-image pages, while \emph{Full Dedup} provides an unsafe lower bound. 

\subsubsection{Analysis of Snapshot Data Transfers}
\label{sec:eval:data_transfers}

\begin{figure}
    \centering
    \includegraphics[width=\linewidth]{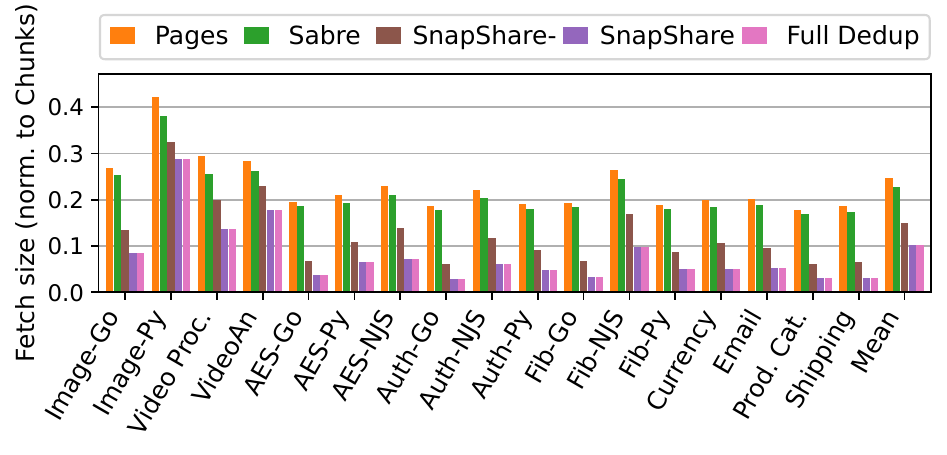}
    \vspace{-25pt}
    \caption{
    Normalized network fetch sizes required for cold-start restoration across 17 serverless benchmarks. Values are normalized against the \emph{Chunks} baseline (not shown). \sys achieves network efficiency comparable to aggressive software compression purely through structural deduplication. Lower is better. 
    % \kev{SplitShare on the fig}
    % \TODO{Normalize to Chunks instead, and do not show them}
    % \leo{Increase font size}
    }
    \vspace{-10pt}
    \label{fig:download_sizes}
\end{figure}

Figure~\ref{fig:download_sizes} shows the volume of snapshot data transferred during a cold start, normalized to \emph{Chunks}. \emph{Pages} and \emph{Sabre} reduce the mean fetch size to 25\% and 23\%, respectively. Structural deduplication reduces the mean further to 15\% for \sys- and 10\% for \sys, within 0.5\% of unsafe \emph{Full Dedup}.

\begin{comment}
\subsubsection{Analysis of Snapshot Data Transfers}

\begin{figure}
    \centering
    \includegraphics[width=\linewidth]{pics/download_sizes.pdf}
    \caption{
    Per-benchmark fetch sizes required for restoration. Numbers are normalized to WS-based retrievals. 
    \TODO{Normalize to Chunks instead, and do not show them}
    %\TODO{prettify, make more readable (relative sizes?), more systems}
    % Numbers are relative to the chunk-based system. REAP gives 10$\times$ reduction, \sys and Sabre give $\sim$20$\times$.
    % \leo{ETA: updated numbers 28th}
    }
    \label{fig:download_sizes}
\end{figure}

Goal: to show the reason for the improvement in the download time.

Observation: chunk-based system creates 3.5-6.5$\times$ more data to fetch compared to page-based and WS. ZipWS can compress the WS by 2.5-4$\times$ creating 3$\times$ reduction in fetch size. \sys reduces the download size by 2$\times$ over WS.

Conclusion: ws and \sys reduce the required retrieval size, leading to proportional improvement in the restoration durations. 
\end{comment}

\subsection{Sensitivity and Efficiency Analysis}
\label{sec:eval:sensitivity}

% \TODO{Add lead-in}

To assess \sys's broader efficiency and robustness, we analyze its performance under constrained network bandwidth and validate the accuracy of the working set estimation. Furthermore, we demonstrate the complementary benefits of integrating state-of-the-art snapshot compression.

\subsubsection{Network Bandwidth Sensitivity}
\label{sec:eval:net_sensi}

\begin{figure}
    \centering
    \includegraphics[width=\linewidth]{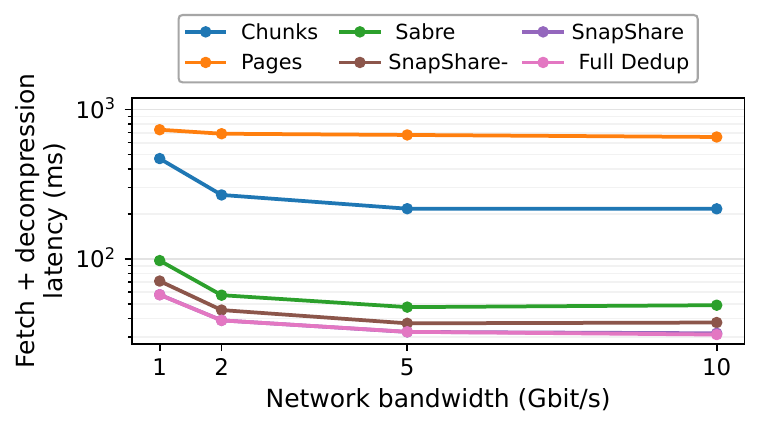}
    \vspace{-20pt}
    \caption{
    Average cold-start fetch time across 17 benchmarks under varying network bandwidth limits. \sys degrades the slowest as network bandwidth decreases, matching the unsafe Full-dedup baseline across all bandwidth settings. Lower is better.
    % \leo{the numbers for chunks don't make sense. with 2gbps it is 4x faster than 1gbps}\lik{Yes, all the other workloads show around a 2× improvement, while video-analytics is an outlier with a 15× improvement. Will check and update.}\lik{Updated.}
    }
    \vspace{-5pt}
    \label{fig:bw_sensitivity}
\end{figure}

%\cz{How does the chunk size change when the network bandwidth changes?}
To evaluate resilience under network constraints, we throttle the bandwidth available for snapshot downloads from 10~Gbps down to 1~Gbps.
% \TODO{check phrasing above}
Figure~\ref{fig:bw_sensitivity} plots average fetch-plus-decompression latency across the benchmark suite. \emph{Pages} benefits little from additional bandwidth because its many small transfers underutilize the link. At 10~Gbps, \sys completes this stage in 31.8~ms, a 6.8$\times$ improvement over \emph{Chunks}. At 1~Gbps, \sys requires 57.7~ms compared with 469.3~ms for \emph{Chunks}. Across the sweep, \sys remains close to unsafe \emph{Full Dedup}; the 12--19\% gap between \sys- and \sys (not visible in the figure) again isolates the benefit of approved container-image sharing.
% \dmi{I suggest moving from 1gb to 10gb point, which is used everywhere else and spelling out that it is as a saturation point for all the configs incl \sys} \leo{i still want to spell out the benefits of \sys under very constrained network, that's the point of this subsection, after all}
% \leo{All systems have limited fetch time improvement with network bandwidth above 10~Gbps, where \sys reduces the fetch time to 12~ms, a 10$\times$ improvement over \emph{Chunks}.} \TODO{reorder, 10gbps saturation -> good even at 1gbps}
% \TODO{add limits for infra in providers to backgrounds}
\sys reduces the raw data volume required for restoration, so operators can confidently allocate the vast majority of the cluster network bandwidth to active user workloads rather than to infrastructure provisioning, e.g., leaving 90\% of a typically deployed NIC's 100Gbps for application use.

\subsubsection{Compression Algorithm Sensitivity}
\label{sec:eval:compression}

% \TODO{add figure and description}

\begin{figure}
    \centering
    \includegraphics[width=0.80\linewidth]{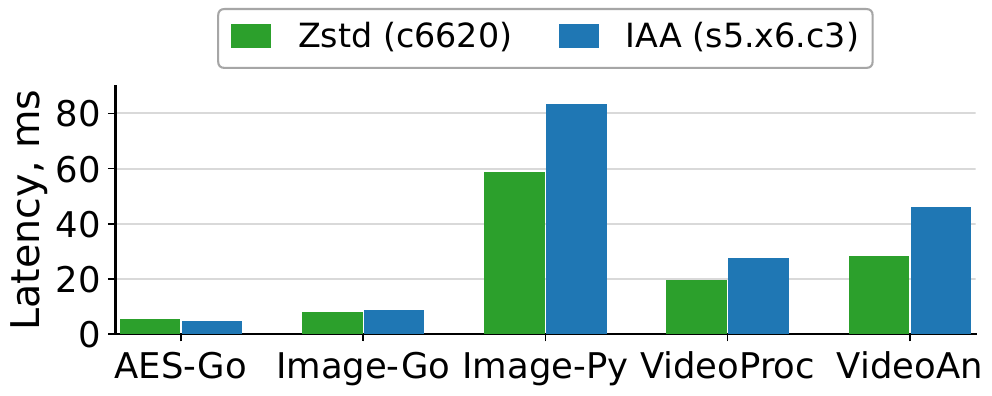}
    \vspace{-10pt}
    \caption{
    % \lik{Decompression latency for identical private working sets across five workloads. Zstd+ achieves lower latency on four of the five workloads.}
    % \leo{please rename in the legend or make the description that would explain those}
    Decompression latency for identical private working sets across five representative benchmarks, comparing our software \emph{zstd} configuration (eight parallel decoder threads) against Sabre's Intel IAA accelerator (default single-queue configuration). Zstd matches or lowers latency on four of the five workloads. Lower is better.
    }
    \vspace{-5pt}
    \label{fig:compression_zstd3_vs_iaa}
\end{figure}

% Zstd+ uses eight parallel decoders, while IAA uses the default setting. 

% s by exploiting greater CPU parallelism, at the cost of higher CPU resource usage. Lower is better

% This supports the 

% \leo{Takeaway: zstd and HW-accelerated both can be hidden behind the fetch latency due to their speed and streamability.}

% We want to verify that zstd is a representative choice.

% Figure~\ref{fig:compression_zstd3_vs_iaa} shows that the latency of zstd decompression used in our evaluation is up to 25\% lower than the IAA-accelerated decompression. This makes zstd representative of state-of-the-art compression in terms of raw decompression performance~\cite{lazarev:sabre}.

Throughout this evaluation, the \emph{Sabre} baseline applies streaming software \emph{zstd} compression rather than reproducing Sabre's original Intel In-Memory Analytics Accelerator (IAA) hardware pipeline~\cite{lazarev:sabre}. Before relying on this substitution, we check that it does not put \emph{Sabre} at an unfair disadvantage in our comparisons, i.e., that zstd is a representative proxy for the decompression latency Sabre's design would achieve.

On the IAA-equipped server described in \S\ref{sec:methodology}, we measure decompression latency for identical private working sets under our zstd configuration (eight parallel decoder threads, as used throughout the rest of the evaluation) and under Sabre's IAA accelerator in its default, single-queue configuration. Figure~\ref{fig:compression_zstd3_vs_iaa} shows that zstd matches or achieves up to 25\% lower decompression latency than IAA on four of the five workloads. Software zstd is therefore at least as fast as Sabre's original hardware-accelerated pipeline in raw decompression latency, confirming that the improvements we report over \emph{Sabre} elsewhere in this evaluation (\S\ref{sec:eval:latencies}, \S\ref{sec:eval:net_sensi}) do not stem from replacing its compression stage with a software-based substitute.

\subsection{Summary and Discussion}
\label{sec:eval:summary}

\sys resolves the tension between rapid instantiation and strict tenant isolation. Selective global deduplication reduces remote-storage requirements by 39\% and local-cache requirements by 38\% relative to \emph{Sabre}. It reduces mean network retrieval volume by 56\% and improves end-to-end tail slowdown by 10\% over \emph{Sabre} and 29\% over \emph{Chunks}.

Unlocking these efficiencies requires acknowledging operational boundaries. While \sys's offline profiling effectively coalesces working sets for idiomatic microservices, highly dynamic monoliths (e.g., JVM frameworks) may miss code paths during deployment-time probes, increasing on-demand remote fetches. This fragility can be mitigated in production via adaptive asynchronous profiling~\cite{aws-snapstart-optimization}, which samples live page faults to seamlessly update working sets in the background without penalizing the critical path.

Additionally, maximizing cache density necessitates a nuanced threat model. \sys structurally isolates sensitive tenant data but accepts execution metadata leakage by permitting timing channels on globally deduplicated, non-sensitive infrastructure (e.g., the guest OS). While an adversary could in theory infer that a co-located victim is using a non-sensitive infrastructure component, this trade-off is necessary for microsecond-scale efficiency and mirrors the industry-accepted standard of deduplicating public container layers in multi-tenant public clouds~\cite{brooker:on-demand}.

\section{Related Works}
% \leo{maybe add more on the memory security in cloud?}
% \TODO{need to update the lead-in}
Extensive research has focused on reducing serverless cold-starts. This section explores the main areas of prior work, highlighting the gap that \sys addresses.

% \rod{Need to add references in citations.}

\textbf{\textit{Optimizing the Execution Environment:}}
A significant body of work has focused on minimizing serverless cold-starts by re-architecting the underlying execution environment. This includes developing lean, serverless-tailored hypervisors such as Firecracker~\cite{agache:firecracker} and gVisor~\cite{web:gvisor}, which strip away unnecessary OS overhead to reduce boot times. Other approaches leverage specialized unikernels~\cite{hand:unikernels} or lightweight container runtimes like RunD~\cite{li:rund} to further accelerate sandbox initialization. Alternatively, systems like Faasm~\cite{shillaker:faasm} and Hydra~\cite{ivanenko:hydra} introduce concurrent sandboxes that execute multiple invocations within a single process to minimize isolation costs. However, across all of these execution environments, memory snapshots have emerged as the mechanism of choice to bypass application-level initialization and further reduce cold-start latency. While these execution frameworks drastically reduce the time to provision a sandbox, they do not resolve the cache efficiency issues and snapshot fetching overheads associated with managing massive application states. In sum, environment-level improvements are orthogonal but complementary to the secure snapshot deduplication mechanism proposed by \sys.

\textbf{\textit{Optimizing Snapshot Restoration:}}
Another line of work introduces various snapshot optimizations to accelerate state initialization by mitigating I/O bottlenecks during the restore process. Systems like REAP~\cite{ustiugov:benchmarking} coalesce memory working sets to improve spatial locality during fetch operations, while FaaSnap~\cite{ao:faasnap} combines this with on-demand paging. Remote and local fork architectures, such as SEUSS~\cite{cadden:seuss}, Catalyzer~\cite{du:catalyzer}, AFaaS~\cite{chai:fork}, and MITOSIS~\cite{wei:no-provisioned}, similarly deliver memory pages on-demand to overlap function execution with data transfer.
Additionally, Sabre~\cite{lazarev:sabre} further optimizes the data movement pipeline through hardware-accelerated decompression, which we used in our evaluation~\S\ref{sec:eval:compression}.
Crucially, to accelerate restoration, these systems inherently rely on duplicated state—either by maintaining warm instances for forking or by storing per-function working sets. \sys can directly improve the memory or storage requirements of such systems by deduplicating the common infrastructure without sacrificing sensitive state isolation.
% Crucially, while \sys can directly leverage these local data movement optimizations, existing solutions operate under simplified security assumptions, \dmi{i.e., they consider the entire snapshot contents as safe}. \dmi{In contrast, \sys applies the production-standards~\cite{brooker:on-demand} of data protection via converged encryption and explicit data provenance.}
% \sout{While these techniques optimize snapshot chunk retrieval, applying them without strict data provenance is unsafe in zero-trust multi-tenant environments.}
% TODO - add Spice (Taming Serverless Cold Starts Through OS Co-Design).

\textbf{\textit{Snapshot Deduplication and Security:}}
A third line of work optimizes the remote storage layer by deduplicating snapshot data. Because many serverless functions share common base images, language runtimes, and libraries, their memory snapshots contain a large amount of redundant data. Systems such as SnapStore~\cite{panda:snapstore}, SnapBPF~\cite{psomadakis:snapbpf}, and TrEnv~\cite{huang:trenv} capitalize on this redundancy by implementing chunk-based deduplication to minimize the storage footprint in the remote registry, enabling worker nodes to sparsely retrieve only the necessary memory pages in a single, highly efficient network I/O operation. However, traditional content-based deduplication is fundamentally unsafe in zero-trust, multi-tenant environments: memory deduplication can introduce cross-domain side channels that let a co-located tenant confirm the presence of specific content~\cite{suzaki:memory-disclosure,xiao:memory-dedup-security,bosman:dedup}. Accordingly, production systems such as Firecracker recommend disabling memory deduplication when strong tenant isolation is required~\cite{firecracker_docs}. Other approaches retain deduplication while limiting cross-domain leakage, e.g., VUsion~\cite{oliverio:vusion} and Slicedup~\cite{vano:slicedup}, while later work revisits the security of same-domain deduplication~\cite{costi:same-domain-dedup}. Closer to our setting, CPC~\cite{lim:cpc} coordinates a shared page cache across serverless functions, and UPM~\cite{qiu:user-guided-dedup} propose user-guided page merging for serverless memory deduplication; both rely on functions or tenants to supply hints about which pages are safe to share. \sys addresses this gap directly: it permits deduplication only when a page's complete content exactly matches a page in a provider-approved source---without requiring tenant hints or attempting to infer historical provenance from the page content---thereby preventing private snapshot pages from ever being shared across tenants.

\section{Conclusion}

% \sys selectively deduplicates \leo{only snapshot pages whose complete contents exactly match provider-approved infrastructure sources}, while strictly isolating all non-matching, tenant-specific private data.
% \leo{This conservative content-match rule classifies modified container-image pages as private rather than relying on inferred provenance.}
% This allows for high cache density and storage deduplication without sacrificing sensitive data isolation. Our evaluation shows that \sys delivers substantial performance benefits: it reduces cold-start retrieval latency by up to 10x compared to chunk-based systems like AWS SnapStart. Additionally, it improves storage and cache efficiency by 40\% to 50\% matching the performance of insecure full-deduplication systems.

\sys selectively deduplicates only snapshot pages whose complete contents exactly match provider-approved infrastructure sources, while strictly isolating all other, tenant-specific data as private. This conservative content-match rule enables high cache density and storage deduplication without sacrificing sensitive data isolation. Implemented as an extension to Knative, \sys reduces remote-storage requirements by 39\% and local-cache footprints by 38\% relative to the tenant-isolated Sabre baseline, closely matching the efficiency of an idealized unsafe full-deduplication design (within 6\% and 3\% for storage and cache, respectively). These savings translate into a 6.8$\times$ reduction in cold-start fetch latency over chunk-based systems like AWS SnapStart, and a 10--29\% reduction in end-to-end tail request slowdown relative to safe baselines.

% USENIX program committees give extra points to submissions that are
% backed by artifacts that are publicly available. If you made your code
% or data available, it's worth mentioning this fact in a dedicated
% section.

%-------------------------------------------------------------------------------

\bibliographystyle{ACM-Reference-Format}
% \bibliography{ref}
\bibliography{gen-abbrev,ref,dblp}

%%%%%%%%%%%%%%%%%%%%%%%%%%%%%%%%%%%%%%%%%%%%%%%%%%%%%%%%%%%%%%%%%%%%%%%%%%%%%%%%
\end{document}